\documentclass[pre,twocolumn,aps,showpacs]{revtex4}
\newcommand{\bec}[1]{\mbox{\boldmath $ #1$}}
\usepackage{graphicx}
\begin{document}
\bigskip
\bigskip
\title{Magnetic fluctuations and formation of large-scale inhomogeneous magnetic
structures in a turbulent convection}
\author{Igor Rogachevskii}
\email{gary@bgu.ac.il} \homepage{http://www.bgu.ac.il/~gary}
\author{Nathan Kleeorin}
\email{nat@bgu.ac.il}
\affiliation{Department of Mechanical
Engineering, The Ben-Gurion
University of the Negev, \\
POB 653, Beer-Sheva 84105, Israel}
\date{\today}
\begin{abstract}
Magnetic fluctuations generated by a tangling of the mean magnetic
field by velocity fluctuations are studied in a developed turbulent
convection with large magnetic Reynolds numbers. We show that the
energy of magnetic fluctuations depends on magnetic Reynolds number
only when the mean magnetic field is smaller than $B_{\rm eq} / 4 {\rm
Rm}^{1/4}$, where $B_{\rm eq}$ is the equipartition mean magnetic
field determined by the turbulent kinetic energy and ${\rm Rm}$ is
magnetic Reynolds number. Generation of magnetic fluctuations in a
turbulent convection with a nonzero mean magnetic field results in a
decrease of the total turbulent pressure and may cause formation of
the large-scale inhomogeneous magnetic structures even in an
originally uniform mean magnetic field. This effect is caused by a
negative contribution of the turbulent convection to the effective
mean Lorentz force. The inhomogeneous large-scale magnetic fields
are formed due to the excitation of the large-scale instability. The
energy for this instability is supplied by the small-scale turbulent
convection. The discussed effects might be useful for understanding
the origin of the solar nonuniform magnetic fields, e.g., sunspots.
\end{abstract}

\pacs{47.65.Md}

\maketitle

\section{Introduction}

  Magnetic fields in astrophysics are strongly nonuniform (see,
e.g., \cite{M78,P79,KR80,ZRS83,RSS88,RH04,O03,BS05}). Large-scale
magnetic structures are observed in the form of sunspots, solar
coronal magnetic loops, etc. There are different mechanisms for the
formation of the large-scale magnetic structures, e.g., the magnetic
buoyancy instability of stratified continuous magnetic field
\cite{P79,P66,G70,P82}, the magnetic flux expulsion \cite{W66}, the
topological magnetic pumping \cite{DY74}, etc.

Magnetic buoyancy applies in the literature for different situations
(see \cite{P82}). The first corresponds to the magnetic buoyancy
instability of stratified continuous magnetic field (see, e.g.,
\cite{P79,P66,G70,P82}), and magnetic flux tube concept is not used
there. The magnetic buoyancy instability of stratified continuous
magnetic field is excited when the scale of variations of the
initial magnetic field is less than the density stratification
length. On the other hand, buoyancy of discrete magnetic flux tubes
has been discussed in a number of studies in solar physics and
astrophysics (see, e.g., \cite{P82,P55,S81,SB82,FS93,SC94}). This
phenomenon is also related to the problem of the storage of magnetic
fields in the overshoot layer near the bottom of the solar
convective zone (see, e.g., \cite{SW80,T01,TH04,B05}).

A universal mechanism of the formation of the nonuniform
distribution of magnetic flux is associated with a magnetic flux
expulsion. In particular, the expulsion of magnetic flux from
two-dimensional flows (a single vortex and a grid of vortices) was
demonstrated in \cite{W66}. In the context of solar and stellar
convection, the topological asymmetry of stationary thermal
convection plays very important role in the magnetic field dynamics.
In particular, the topological magnetic pumping is caused by the
topological asymmetry of the thermal convection \cite{DY74}. The
fluid rises at the centers of the convective cells and falls at
their peripheries. The ascending fluid elements (contrary to the
descending fluid elements) are disconnected from one another. This
causes a topological magnetic pumping effect allowing downward
transport of the mean horizontal magnetic field to the bottom of a
cell but impeding its upward return \cite{ZRS83,DY74,GP83}.

Turbulence may form inhomogeneous large-scale magnetic fields due to
turbulent diamagnetic and paramagnetic effects (see, e.g.,
\cite{KR80,Z57,VK83,K91,KR92,RKR03}). Inhomogeneous velocity
fluctuations lead to a transport of mean magnetic flux from regions
with high intensity of the velocity fluctuations. Inhomogeneous
magnetic fluctuations due to the small-scale dynamo cause turbulent
paramagnetic velocity, i.e., the magnetic flux is pushed into
regions with high intensity of the magnetic fluctuations. Another
effects are the effective drift velocities of the mean magnetic
field caused by inhomogeneities of the fluid density \cite{K91,KR92}
and pressure \cite{KP93}. In a nonlinear stage of the magnetic field
evolution, inhomogeneities of the mean magnetic field contribute to
the diamagnetic or paramagnetic drift velocities depending on the
level of magnetic fluctuations due to the small-scale dynamo and
level of the mean magnetic field \cite{RK04}. The diamagnetic
velocity causes a drift of the magnetic field components from the
regions with a high intensity of the mean magnetic field.

The nonlinear drift velocities of the mean magnetic field in a
turbulent convection have been determined in \cite{RK06}. This study
demonstrates that the nonlinear drift velocities are caused by the
three kinds of the inhomogeneities, i.e., inhomogeneous turbulence;
the nonuniform fluid density and the nonuniform turbulent heat flux.
The nonlinear drift velocities of the mean magnetic field cause the
small-scale magnetic buoyancy and magnetic pumping effects in the
turbulent convection. These phenomena are different from the
large-scale magnetic buoyancy and magnetic pumping effects which are
due to the effect of the mean magnetic field on the large-scale
density stratified fluid flow. The small-scale magnetic buoyancy and
magnetic pumping can be stronger than these large-scale effects when
the mean magnetic field is smaller than the equipartition field
determined by the turbulent kinetic energy \cite{RK06}. The pumping
of magnetic flux in three-dimensional compressible magnetoconvection
has been studied in direct numerical simulations in \cite{OS02} by
calculating the turbulent diamagnetic and paramagnetic velocities.

Turbulence may affect also the Lorentz force of the large-scale
magnetic field (see \cite{KRR89,KRR90,KR94,KMR96}). This effect can
also form inhomogeneous magnetic structures. In this study a
theoretical approach proposed in \cite{KRR89,KRR90,KR94,KMR96} for a
nonconvective turbulence is further developed and applied to
investigate the modification of the large-scale magnetic force by
turbulent convection and to elucidate a mechanism of formation of
inhomogeneous magnetic structures.

This paper is organized as follows. In Sect.~II we discuss the
physics of the effect of turbulence on the large-scale Lorentz
force. In Sect.~III we formulate the governing equations, the
assumptions, the procedure of the derivations of the large-scale
effective magnetic force in turbulent convection. In Sect.~IV we
study magnetic fluctuations and determine the modification of the
large-scale effective Lorentz force by the turbulent convection. In
Sect.~V we discuss  formation of the large-scale magnetic
inhomogeneous structures in the turbulent convection due to
excitation of the large-scale instability. Finally, we draw
conclusions in Sect.~VI. In Appendix~A we perform the derivation of
the large-scale effective Lorentz force in the turbulent convection.

\section{Turbulent pressure and effective mean magnetic pressure}

In this Section we discuss the physics of the effect of turbulence
on the large-scale Lorentz force. First, let us examine an isotropic
turbulence. The Lorentz force of the small-scale magnetic
fluctuations can be written in the form ${\cal F}_i^{(m)} =
\nabla_{j} \sigma_{ij}^{(m)}$, where the magnetic stress tensor
$\sigma_{ij}^{(m)}$ is given by
\begin{eqnarray}
\sigma_{ij}^{(m)} = - {\langle {\bf b}^2 \rangle \over 2} \,
\delta_{ij} + \langle b_i b_j \rangle \;, \label{I1}
\end{eqnarray}
${\bf b}$ are the  magnetic fluctuations and $ \delta_{ij} $ is the
Kronecker tensor. Hereafter we omit the magnetic permeability of the
fluid $\mu$ and include $\mu^{-1/2}$ in the definition of magnetic
field. The angular brackets in Eq.~(\ref{I1}) denote the ensemble
averaging. For isotropic turbulence $\langle b_i b_j \rangle =
\delta_{ij} \, \langle {\bf b}^2 \rangle / 3 $, and the magnetic
stress tensor reads
\begin{eqnarray}
\sigma_{ij}^{(m)} = - {\langle {\bf b}^2 \rangle \over 6} \,
\delta_{ij} = - {W_m \over 3} \, \delta_{ij} \;,
\label{I2}
\end{eqnarray}
where $W_m = \langle {\bf b}^2 \rangle/ 2$ is the energy density of
the magnetic fluctuations. The magnetic pressure $p_m$ is related to
the magnetic stress tensor: $\sigma_{ij}^{(m)} = - p_m \,
\delta_{ij}$, where $p_m = W_m / 3$. Similarly, in an isotropic
turbulence the Reynolds stresses $\langle u_i u_j \rangle$ read:
$\langle u_i u_j \rangle = \delta_{ij} \langle {\bf u}^2 \rangle /
3$, and
\begin{eqnarray}
\sigma_{ij}^{(v)} = - \rho_0 \langle u_i u_j \rangle = - {\rho_0
\langle {\bf u}^2 \rangle \over 3} \, \delta_{ij} = - {2 W_k \over
3} \, \delta_{ij} \;,
\label{I3}
\end{eqnarray}
where ${\bf u}$ are the velocity fluctuations, $W_k = \rho_0 \langle
{\bf u}^2 \rangle/ 2$ is the kinetic energy density of the velocity
fluctuations and $\rho_0$ is the fluid density. Equation~(\ref{I3})
yields the hydrodynamic pressure $p_v = 2 W_k / 3$, where
$\sigma_{ij}^{(v)} = - p_v \, \delta_{ij}$. Therefore, the equation
of state for the isotropic turbulence is given by
\begin{eqnarray}
p_{_{T}} = \frac{1}{3} W_m + \frac{2}{3} W_k \;,
\label{WA1}
\end{eqnarray}
(see also \cite{LL75,LL84}), where $p_{_{T}}$ is the total
(hydrodynamic plus magnetic) turbulent pressure. Similarly, the
equation of state for an anisotropic turbulence reads
\begin{eqnarray}
p_{_{T}} = \frac{2}{3(2 + A_N)} W_m + {4 + 3\, A_N \over 3(2 + A_N)}
\, W_k \;, \label{RWA1}
\end{eqnarray}
where $A_N = (2/3)
[\langle {\bf u}_{\perp}^{2} \rangle / \langle {\bf u}_{z}^{2}
\rangle - 2]$ is the degree of anisotropy of the turbulent
velocity field ${\bf u} = {\bf u}_{\perp} + u_{z} {\bf e}$.
For an isotropic three-dimensional turbulence $\langle {\bf u}_{\perp}^{2}
\rangle = 2 \langle {\bf u}_{z}^{2}\rangle$ and the parameter $A_N = 0$, while for a two-dimensional turbulence $\langle {\bf u}_{z}^{2}\rangle=0$ and the
degree of anisotropy $A_N \to \infty$. Here ${\bf e}$ is the
vertical unit vector perpendicular to the plane of the
two-dimensional turbulence.

In a two-dimensional turbulence $A_N \to \infty$ and the total turbulent pressure $p_{_{T}} \to W_k$. Note that the magnetic pressure in a
two-dimensional turbulence vanishes. Indeed, for isotropic magnetic fluctuations
in a two-dimensional turbulence $\langle b_i b_j \rangle= (1/2) \,
\langle {\bf b}^2 \rangle \, \delta_{ij}^{(2)}$, and therefore,
$\sigma_{ij}^{(m)} \equiv - (1/2) \, \langle {\bf b}^2 \rangle \,
\delta_{ij}^{(2)} + \langle b_i b_j \rangle = 0$, where
$\delta_{ij}^{(2)}=\delta_{ij}-e_i\, e_j$.

The total energy density $ W_T = W_k + W_m $ of the homogeneous
turbulence with a mean magnetic field ${\bf B}$ is determined by the
equation
\begin{eqnarray}
\frac{\partial W_T}{ \partial t} = I_T - \frac{W_T}{\tau_0} +
\eta_{_{T}} \, (\bec{\nabla} \times {\bf B})^2  \;, \label{WA3}
\end{eqnarray}
(see, e.g., \cite{KMR96}), where $\tau_0$ is the  correlation time
of the turbulent velocity field in the maximum scale $l_0$ of
turbulent motions, $I_T$ is the energy source of turbulence,
$\eta_{_{T}}$ is the turbulent magnetic diffusion and the mean
magnetic field ${\bf B}$ is given (prescribed). The second term,
$W_T / \tau_0$, in the right hand side of Eq.~(\ref{WA3}) determines
the dissipation of the turbulent energy. For a given
time-independent source of turbulence $I_T$ the solution of
Eq.~(\ref{WA3}) is given by
\begin{eqnarray}
W_T &=& \tau_0 \, \big[I_T + \eta_{_{T}} \, (\bec{\nabla} \times
{\bf B})^2 \big] \, \Big[1 - \exp\Big(-{t \over \tau_0} \Big) \Big]
\nonumber\\
&& + \tilde{W}_T \exp\Big(-{t \over \tau_0} \Big) \;, \label{WAA3}
\end{eqnarray}
where $\tilde{W}_T = W_T(t=0)$. For instance, a time-independent
source of the turbulence exists in the Sun. The mean nonuniform
magnetic field causes an additional energy source of the turbulence,
$I_N = \eta_{_{T}} (\bec{\nabla} \times {\bf B})^2$. The ratio $I_N/
I_T$ of these two sources of turbulence is of the order of
\begin{eqnarray}
{I_N \over I_T} \simeq \biggl({l_0 \over L_B}\biggr)^2 {{\bf B}^2
\over \rho_0 \langle {\bf u}^2 \rangle^{(0)}} \ll 1 \;,
\label{I4}
\end{eqnarray}
where $L_B$ is the characteristic scale of the spatial variations of
the mean magnetic field. Since $ l_0 \ll L_B $ and ${\bf B}^2 \ll
\rho_0 \langle {\bf u}^2 \rangle^{(0)}$, we can neglect the small
magnetic source $I_N$ of the turbulence.  Thus, for $t \gg \tau_0$
the total energy density of the turbulence reaches a steady state
$W_T = const = \tau_0 \, I_T$. Therefore, the total energy density
$W_T$ of the homogeneous turbulence is conserved (the dissipation is
compensated by a supply of energy), i.e
\begin{eqnarray}
W_k + W_m = const . \label{WA2}
\end{eqnarray}
A more rigorous derivation of Eq.~(\ref{WA2}) is given in Appendix A
(see Eq.~(\ref{B27})). Equation~(\ref{WA2}) implies that the uniform
large-scale magnetic field performs no work on the turbulence. It
can only redistribute the energy between hydrodynamic and magnetic
fluctuations.

Combining Eqs.~(\ref{WA1}) and~(\ref{WA2}) we can express the change
of turbulent pressure $\delta p_{_{T}}$ in terms of the change of
the magnetic  energy density $\delta W_m$ for an isotropic
turbulence $ \delta p_{_{T}} = - (1 / 3) \,  \delta W_m$ (see
\cite{KRR89,KRR90,KMR96}). Therefore, the turbulent pressure is
reduced when magnetic fluctuations are generated ($\delta W_m
> 0$). Similarly, for an anisotropic turbulence, the generation of
magnetic fluctuations reduces the turbulent pressure, i.e.,
\begin{eqnarray*}
\delta p_{_{T}} = - {2 + 3\, A_N \over 3(2 + A_N)} \delta W_m \; .
\label{WA4}
\end{eqnarray*}

The total turbulent pressure is decreased also by the tangling of
the large-scale mean magnetic field ${\bf B}$ by the velocity
fluctuations (see, e.g., \cite{M78,P79,KR80,ZRS83}, and references
therein). The mean magnetic field generates additional small-scale
magnetic fluctuations due to a tangling of the mean magnetic field
by velocity fluctuations. For a small energy of the mean magnetic
field, ${\bf B}^2 \ll \rho_0 \, \langle {\bf u}^2 \rangle$, the
energy of magnetic fluctuations, $\langle {\bf b}^2 \rangle - \langle {\bf b}^2 \rangle^{(0)}$, caused by a tangling of the mean magnetic field can be written in the form:
\begin{eqnarray}
\langle {\bf b}^2 \rangle - \langle {\bf b}^2 \rangle^{(0)} = a_m(B,
{\rm Rm}) {\bf B}^2 + O[{\bf B}^4 / (\rho_0 \, \langle {\bf u}^2 \rangle)^2] \;,
\nonumber\\
\label{WAA2}
\end{eqnarray}
where $\langle {\bf b}^2 \rangle^{(0)}$ are the magnetic
fluctuations with a zero mean magnetic field generated by a
small-scale dynamo. Equation~(\ref{WAA2}) allows us to determine the
variation of the magnetic energy $\delta W_m$. Therefore, the total
turbulent pressure reads
\begin{eqnarray}
p_{_{T}} = p_{_{T}}^{(0)} - q_p \, {{\bf B}^2  \over 2} \;,
\label{WA5}
\end{eqnarray}
where $p_{_{T}}^{(0)}$ is the turbulent pressure in a flow with a
zero mean magnetic field and the coefficient $q_p \propto a_m(B,
{\rm Rm})$. Here we neglect the small terms $\sim O[{\bf B}^4 / (\rho_0 \, \langle {\bf u}^2 \rangle)^2]$. The coefficient $q_p$ is positive when magnetic
fluctuations are generated, and it is negative when they are damped.
The total pressure is
\begin{eqnarray}
P_{\rm tot} \equiv P_k + p_{_{T}} + P_B(B) = P_k + p_{_{T}}^{(0)} +
(1- q_p) \, {{\bf B}^2  \over 2} \;,
\nonumber\\
\label{WA7}
\end{eqnarray}
where $P_k$ is the mean fluid pressure and $P_B(B) = {\bf B}^2 / 2$
is the  magnetic pressure of the mean magnetic field. Now we examine
the part of the total pressure $P_{\rm tot}$ that depends on the
mean magnetic field ${\bf B}$, i.e., we consider
\begin{eqnarray}
P_m(B) &=& P_B(B) - q_p(B) \, {{\bf B}^2  \over 2} = (1 - q_p) \,
{{\bf B}^2  \over 2} \;,
\nonumber\\
\label{WA6}
\end{eqnarray}
(see \cite{KRR89,KRR90,KMR96}), where now $P_{\rm tot} = P + P_m(B)$
and $P = P_k + p_{_{T}}^{(0)}$. The pressure $P_m(B)$ is called the
effective (or combined) mean magnetic pressure. Note that both the
hydrodynamic and magnetic fluctuations contribute to the combined
mean magnetic pressure. However, the gain in the turbulent magnetic
pressure $p_m$ is not as large as the reduction of the turbulent
hydrodynamic pressure $p_v$ by the mean magnetic field ${\bf B}$.
This is due to different coefficients  multiplying by $W_m$ and
$W_k$ in the equation of state~(\ref{WA1}) [see also
Eq.~(\ref{RWA1})]. Therefore, this effect is caused by a negative
contribution of the turbulence to the combined mean magnetic
pressure.

We consider the case when $P \gg {\bf B}^2/2$, so that the total
pressure $P_{\rm tot}$ is always positive. Only the combined mean
magnetic pressure $P_m(B)$ may be negative when $q_p >1$, while the
pressure $P_B(B)$ as well as the values $P_k, \, p_v, \, p_m, \,
p_{_{T}}$ are always positive. When a mean magnetic field $ {\bf B}
$ is superimposed on an isotropic turbulence, the isotropy breaks
down. Nevertheless Eq.~(\ref{WA6}) remains valid, while the
relationship between $q_p$ and $a_m$ may change.

In this Section we use the conservation law~(\ref{WA2}) for the
total turbulent energy only for the elucidation of the principle of
the effect, but we have not employed Eq.~(\ref{WA2}) to develop the
theory of this effect (see for details \cite{KRR90,KR94,KMR96}). In
particular, the high-order closure procedure \cite{KRR90,KMR96} and
the renormalization procedure \cite{KR94} have been used for the
investigation of the nonconvective turbulence at large magnetic and
hydrodynamic Reynolds numbers.

In this study we investigate the modification of the large-scale
magnetic force by turbulent convection. We demonstrate that the
turbulent convection enhances modification of the effective magnetic
force and causes a large-scale instability. This results in
formation of the large-scale inhomogeneous magnetic structures.

\section{Governing equations and the procedure of derivation}

In order to study magnetic fluctuations and the modification of the
large-scale Lorentz force by turbulent convection we use a mean
field approach in which the magnetic and velocity fields, and
entropy are decomposed into the mean and fluctuating parts, where
the fluctuating parts have zero mean values. We assume that there
exists a separation of scales, i.e., the maximum scale of turbulent
motions $l_0$ is much smaller then the characteristic scale $L_B$ of
the mean magnetic field variations. We apply here an approach which
is described in \cite{RK04,RK06,RK07} and outlined below.

We consider a nonrotating turbulent convection with large Rayleigh
numbers and large magnetic Reynolds numbers. We use the equations
for fluctuations of the fluid velocity, ${\bf u}$, entropy, $s'$,
and the magnetic field, ${\bf b}$.  The equations for velocity and
entropy fluctuations are rewritten in the new variables ${\bf v} =
\sqrt{\rho_0} \, {\bf u}$ and $s = \sqrt{\rho_0} \, s'$. We also use
the new variable ${\bf H} = {\bf B} / \sqrt{\rho_0}$ for the mean
magnetic field, ${\bf B}$. On the other hand, we do not use a new
variable for magnetic fluctuations, ${\bf b}$. Equations for
fluctuations of fluid velocity, entropy and magnetic field are
applied in the anelastic approximation, that is a combination of the
Boussinesq approximation and the condition ${\rm div} \, (\rho_0 \,
{\bf u}) = 0$. The turbulent convection is regarded as a small
deviation from a well-mixed adiabatic reference state. This implies
that we consider the hydrostatic nearly isentropic basic reference
state.

Using these equations for fluctuations of fluid velocity, entropy
and magnetic field written in a Fourier space we derive equations
for the two-point second-order correlation functions of the velocity
fluctuations $\langle v_i \, v_j\rangle$, the magnetic fluctuations
$\langle b_i \, b_j \rangle$, the entropy fluctuations $\langle s \,
s\rangle$, the cross-helicity $\langle b_i \, v_j \rangle$, the
turbulent heat flux $\langle s \, v_i \rangle$ and $\langle s \, b_i
\rangle$. The equations for these correlation functions are given by
Eqs.~(\ref{B6})-(\ref{B11}) in Appendix A. We split the tensor
$\langle b_i \, b_j \rangle$ of magnetic fluctuations into
nonhelical, $h_{ij}$, and helical, $h_{ij}^{(H)},$ parts. The
helical part $h_{ij}^{(H)}$ depends on the magnetic helicity (see
below). We also split all second-order correlation functions,
$M^{(II)}$, into symmetric and antisymmetric parts with respect to
the wave vector ${\bf k}$, e.g., $h_{ij} = h_{ij}^{(s)} +
h_{ij}^{(a)}$, where the tensors $h_{ij}^{(s)} = [h_{ij}({\bf k}) +
h_{ij}(-{\bf k})] / 2$ describes the symmetric part of the tensor
and $h_{ij}^{(a)} = [h_{ij}({\bf k}) - h_{ij}(-{\bf k})] / 2$
determines the antisymmetric part of the tensor.

The second-moment equations include the first-order spatial
differential operators $\hat{\cal N}$  applied to the third-order
moments $M^{(III)}$. A problem arises how to close the equations for
the second moments, i.e., how to express the third-order terms
$\hat{\cal N} M^{(III)}$ through the second moments $M^{(II)}$ (see,
e.g., \cite{O70,MY75,Mc90}). We will use the spectral $\tau$
approximation which postulates that the deviations of the
third-moment terms, $\hat{\cal N} M^{(III)}({\bf k})$, from the
contributions to these terms afforded by the background turbulent
convection, $\hat{\cal N} M^{(III,0)}({\bf k})$, are expressed
through the similar deviations of the second moments, $M^{(II)}({\bf
k}) - M^{(II,0)}({\bf k})$:
\begin{eqnarray}
\hat{\cal N} M^{(III)}({\bf k}) &-& \hat{\cal N} M^{(III,0)}({\bf
k})
\nonumber\\
&=& - {1 \over \tau(k)} \, [M^{(II)}({\bf k}) - M^{(II,0)}({\bf k})]
\;,
\nonumber\\
\label{AAC3}
\end{eqnarray}
(see, e.g., \cite{O70,PFL76,KRR90,KMR96,RK04}), where $\tau(k)$ is
the scale-dependent relaxation time. In the background
turbulent convection the mean magnetic field is zero.
The $\tau$ approximation is applied for large hydrodynamic
and magnetic Reynolds numbers, and large Rayleigh
numbers. In this case there is only one relaxation time $\tau$ which can be
identified with the correlation time of the turbulent velocity
field. A justification of the $\tau$ approximation for different situations
has been performed in numerical simulations and theoretical studies
in \cite{BF02,FB02,BK04,BSM05,SSB07,BS07} (see also review
\cite{BS05}). The $\tau$ approximation is also discussed in Sect.~VI.

We apply the spectral $\tau$ approximation only for the nonhelical
part $h_{ij}$ of the tensor of magnetic fluctuations. The helical
part $h_{ij}^{(H)}$ depends on the magnetic helicity, and it is
determined by the dynamic equation which follows from the magnetic
helicity conservation arguments (see, e.g.,
\cite{KR82,GD94,KR99,KMRS2000,BF2000,BB02,KKMRS03}, review
\cite{BS05} and references therein). The characteristic time of
evolution of the nonhelical part of the tensor $h_{ij}$ is of the
order of the turbulent time $\tau_{0} = l_{0} / u_{0}$, while the
relaxation time of the helical part of the tensor $h_{ij}^{(H)}$ of
magnetic fluctuations is of the order of $\tau_{0} \, {\rm Rm}$,
where ${\rm Rm}= l_0 u_{0} / \eta$ is the magnetic Reynolds number,
$u_{0}$ is the characteristic turbulent velocity in the maximum
scale of turbulent motions $l_{0}$ and $\eta$ is the magnetic
diffusivity due to electrical conductivity of the fluid.

In this study we consider an intermediate nonlinearity which implies that
the mean magnetic field is not strong enough in order to affect the
correlation time of turbulent velocity field.
The theory can be expanded to the case a very strong mean magnetic field
after taking into account a dependence of the correlation time of the turbulent
velocity field on the mean magnetic field.

We assume that the characteristic time of variation of the mean
magnetic field ${\bf B}$ is substantially larger than the
correlation time $\tau(k)$ for all turbulence scales. This allows us
to get a stationary solution for the equations for the second-order
moments given by Eqs.~(\ref{B22})-(\ref{B21B}) in Appendix A. For
the integration in $ {\bf k} $-space of the second moments we have
to specify a model for the background turbulent convection (with a
zero mean magnetic field, $ {\bf B} = 0)$. Here we use the model of
the background shear-free turbulent convection with a given heat
flux (see Eqs.~(\ref{K1})-(\ref{K3}) in Appendix). In this model
velocity and magnetic fluctuations are homogeneous and isotropic.

This procedure allows us to study magnetic fluctuations with a
nonzero mean magnetic field and to investigate the modification of
the large-scale Lorentz force by turbulent convection (see
Sect.~IV).

\section{Magnetic fluctuations and large-scale effective Lorentz force}

\subsection{Magnetic fluctuations with a nonzero mean magnetic field}

Let us study magnetic fluctuations with a nonzero mean magnetic
field using the approach outlined in Sect. III. Integration in ${\bf
k}$ space in Eq.~(\ref{B24}) yields an analytical expression for the
energy of magnetic fluctuations, $\langle {\bf b}^2 \rangle$ [see
Eq.~(\ref{D1}) in Appendix A]. The energy of magnetic fluctuations
versus the mean magnetic field ${B}/{B}_{\rm eq}$ is shown in
Fig.~1, where $B_{\rm eq}$ is the equipartition mean magnetic field
determined by the turbulent kinetic energy. The asymptotic formulae
for $\langle {\bf b}^2 \rangle$ are given below. In particular, for
a very weak mean magnetic field, $B \ll B_{\rm eq} / 4 {\rm
Rm}^{1/4}$, the energy of magnetic fluctuations is given by
\begin{eqnarray}
\langle {\bf b}^2 \rangle &=& \langle {\bf b}^2 \rangle^{(0)} + {4
\over 3} [\langle {\bf v}^2 \rangle^{(0)} - \langle {\bf b}^2
\rangle^{(0)}] \, {B}^2 \, \ln {\rm Rm}
\nonumber \\
&& + {8 a_\ast \over 5} \langle {\bf v}^2 \rangle^{(0)} \, {B}^2 \,
(2 - 3 \cos^2 \phi)   \; .
\label{D2}
\end{eqnarray}
where the quantities with the superscript $(0)$ correspond to the
background turbulent convection (with a zero mean magnetic field),
$\langle {\bf v}^2 \rangle^{(0)}$ and $\langle {\bf b}^2
\rangle^{(0)}$ are the velocity and magnetic fluctuations in the
background turbulent convection. Here the magnetic field $B$ is
measured in the units of $B_{\rm eq}$ and $\phi$ is the angle
between the vertical unit vector ${\bf e}$ and the mean magnetic
field ${\bf B}$. The unit vector ${\bf e}$ is directed opposite to
the gravity field. The parameter $a_\ast$ characterizing the
turbulent convection is determined by the budget equation for the
total energy, and it is given by
\begin{eqnarray*}
a_\ast^{-1} = 1 + { \nu_{_{T}} (\nabla  U)^2 + \eta_{_{T}} (\nabla
 B)^2 /\rho_0 \over g F_\ast} \;,
\end{eqnarray*}
where $\nu_{_{T}}$ is the turbulent viscosity, $U$ is the mean fluid
velocity and $F_\ast = \langle u_z \, s' \rangle^{(0)}$ is the
vertical heat flux in the background turbulent convection. The
energy of magnetic fluctuations for a very weak mean magnetic field,
$B \ll B_{\rm eq} / 4 {\rm Rm}^{1/4}$, depends on the magnetic
Reynolds number: $\langle {\bf b}^2 \rangle \propto \ln {\rm Rm}$.
This is an indication of that the spectrum of magnetic fluctuations
is $k^{-1}$ in the limit of a small yet finite mean magnetic field
\cite{RS82,KRR90,KR94,KMR96} (see also discussion in \cite{SIC07}).
When the mean magnetic field $ { B}_{\rm eq} / 4 {\rm Rm}^{1/4} \ll
B \ll { B}_{\rm eq} / 4$, the energy of magnetic fluctuations is
given by
\begin{eqnarray}
\langle {\bf b}^2 \rangle &=& \langle {\bf b}^2 \rangle^{(0)} + {16
\over 3} [\langle {\bf v}^2 \rangle^{(0)} - \langle {\bf b}^2
\rangle^{(0)}] \, {B}^2 \, |\ln (4{B})|
\nonumber \\
&& + {8 a_\ast \over 5} \langle {\bf v}^2 \rangle^{(0)} \, B^2 \, (2
+ 3 \cos^2 \phi)   \;,
\label{D3}
\end{eqnarray}
and for $B \gg { B}_{\rm eq} / 4$ it is given by
\begin{eqnarray}
\langle {\bf b}^2 \rangle &=& {1 \over 2} [\langle {\bf v}^2
\rangle^{(0)} + \langle {\bf b}^2\rangle^{(0)}] - {\pi \over 24 {B}}
[\langle {\bf v}^2 \rangle^{(0)} - \langle {\bf b}^2 \rangle^{(0)}]
\nonumber \\
&& + {\pi a_\ast \over 40 \, {B}} \langle {\bf v}^2 \rangle^{(0)} \,
(1 - 3 \cos^2 \phi)   \; . \label{D4}
\end{eqnarray}
The normalized energy of magnetic fluctuations $\langle {\bf b}^2
\rangle / {\bf B}^2 $ versus the mean magnetic field is shown in
Fig.~2 for a nonconvective and convective turbulence. Inspection of
Figs.~1-2 shows that turbulent convection increases the level of
magnetic fluctuations in comparison with the nonconvective
turbulence. It follows from Eqs.~(\ref{D2})-(\ref{D4}) that in the
case of Alfv\'{e}nic equipartition, $\langle {\bf u}^2 \rangle^{(0)}
= \langle {\bf b}^2 \rangle^{(0)}$, a deviation of the energy of
magnetic fluctuations from the background level is caused by the
turbulent convection.

\begin{figure}
\centering
\includegraphics[width=8cm]{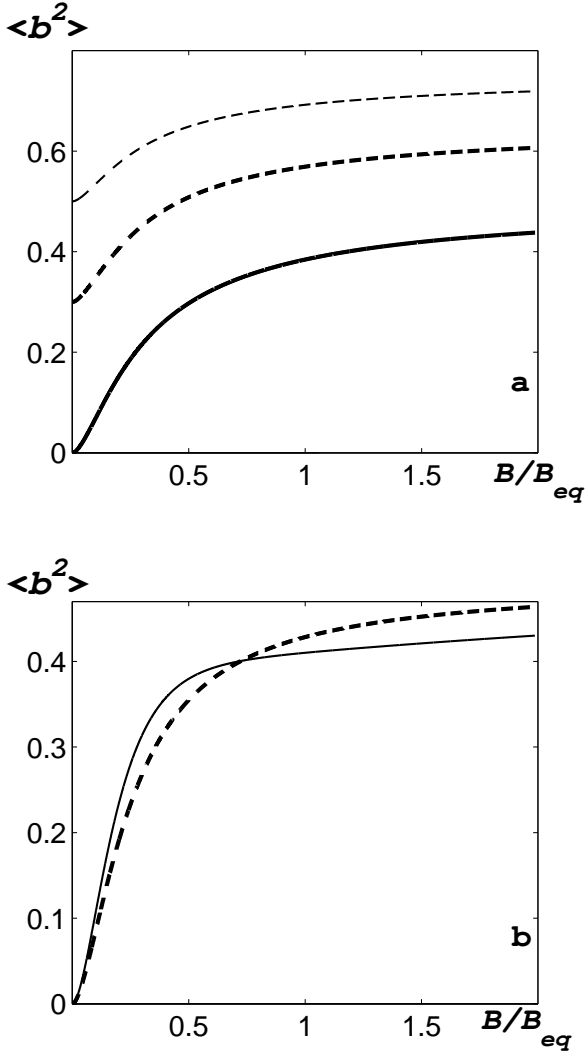}
\caption{\label{Fig1} {\bf (a).} The energy of magnetic fluctuations
$\langle {\bf b}^2 \rangle$ versus the mean magnetic field
${B}/{B}_{\rm eq}$ for a nonconvective turbulence $(a_\ast = 0)$,
$\, \, {\rm Rm}=10^6$ and different values of the parameter
$\epsilon \equiv \langle {\bf b}^2 \rangle^{(0)} / \langle {\bf u}^2
\rangle^{(0)}$: $\, \, \, \epsilon=0$ (solid line); $\epsilon=0.3$
(dashed line) and
$\epsilon=0.5$ (thin dashed). \\
{\bf (b).} The energy of magnetic fluctuations $\langle {\bf b}^2
\rangle$ versus the horizontal (dashed line) and vertical (thin
solid line) mean magnetic field for a convective turbulence $(a_\ast
= 0.7)$, and for ${\rm Rm}=10^6$, $\, \epsilon=0$.}
\end{figure}

\begin{figure}
\centering
\includegraphics[width=8cm]{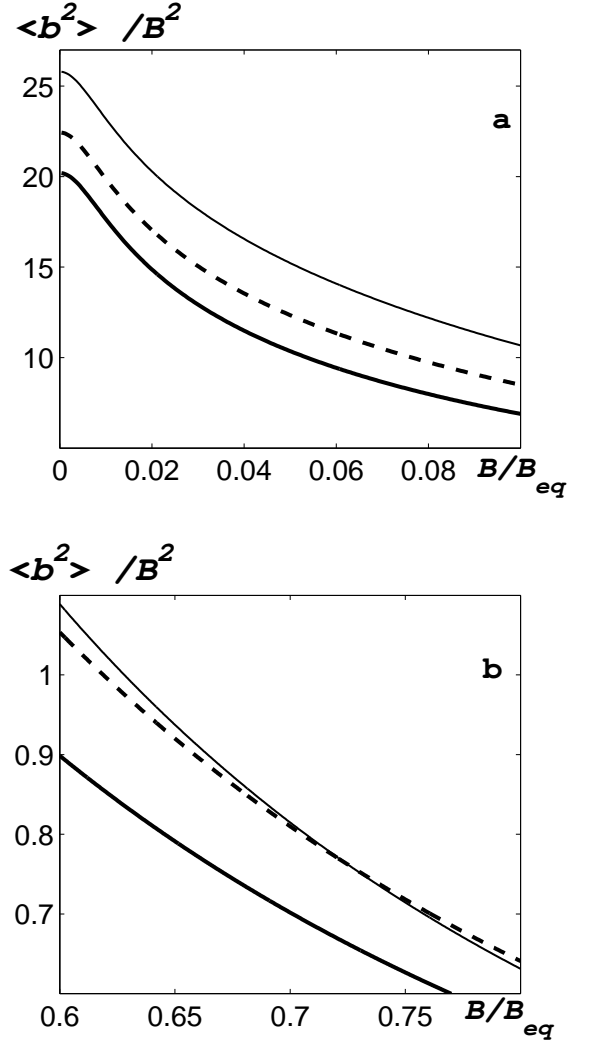}
\caption{\label{Fig2} The normalized energy of magnetic fluctuations
$\langle {\bf b}^2 \rangle / {\bf B}^2 $ versus the mean magnetic
field ${B}/{B}_{\rm eq}$ for a nonconvective turbulence $(a_\ast =
0)$ (solid line); and for a convective turbulence $(a_\ast = 0.7)$
with the horizontal (dashed line) and vertical (thin solid line)
mean magnetic field, where the cases $B \ll B_{\rm eq}$ (Fig. 2a)
and $B \sim B_{\rm eq}$ (Fig. 2b) are shown. Here ${\rm Rm}=10^6$
and $\epsilon=0$.}
\end{figure}

\subsection{The large-scale effective Lorentz force}

The effective (combined) mean magnetic force which takes into
account the effect of turbulence on magnetic force, can be written
in the form ${\cal F}_i^{\rm eff} = \nabla_j \sigma_{ij}^{\rm eff}$,
where the effective stress tensor $\sigma_{ij}^{\rm eff}$ reads
\begin{eqnarray}
\sigma_{ij}^{\rm eff} &=& - {1 \over 2}  \,  {\bf B}^2 \,
\delta_{ij} + B_i  B_j - {1 \over 2} \, \langle {\bf b}^2 \rangle \,
\delta_{ij} + \langle b_i b_j \rangle
\nonumber\\
&& - \rho_0 \, \langle u_i u_j \rangle \; . \label{RWA8}
\end{eqnarray}
The last three terms in RHS of Eq.~(\ref{RWA8}) determine the
contribution of velocity and magnetic fluctuations to the effective
(combined) mean magnetic force. Using Eqs. (\ref{B22})-(\ref{B24})
for $\langle u_i u_j \rangle$ and $\langle b_i b_j \rangle$ after
the integration in ${\bf k}$ space we arrive at the expression for
the effective stress tensor:
\begin{eqnarray}
\sigma_{ij}^{\rm eff} &=& - [1 - q_p(B)] \, {B^2 \over 2} \,
\delta_{ij}  + [1 - q_s(B)] \,  B_i B_j
\nonumber \\
&& + a_\ast \, \sigma_{ij}^A(B) \;, \label{X3}
\end{eqnarray}
where the analytical expressions for the nonlinear coefficients
$q_p(B)$ and $q_s(B)$ are given by Eqs.~(\ref{X1}) and~(\ref{X2}) in
Appendix A, the tensor $\sigma_{ij}^A(B)$ is the anisotropic
contribution caused by turbulent convection to the effective stress
tensor (which is given by Eq.~(\ref{X4}) in Appendix A).

\begin{figure}
\centering
\includegraphics[width=8cm]{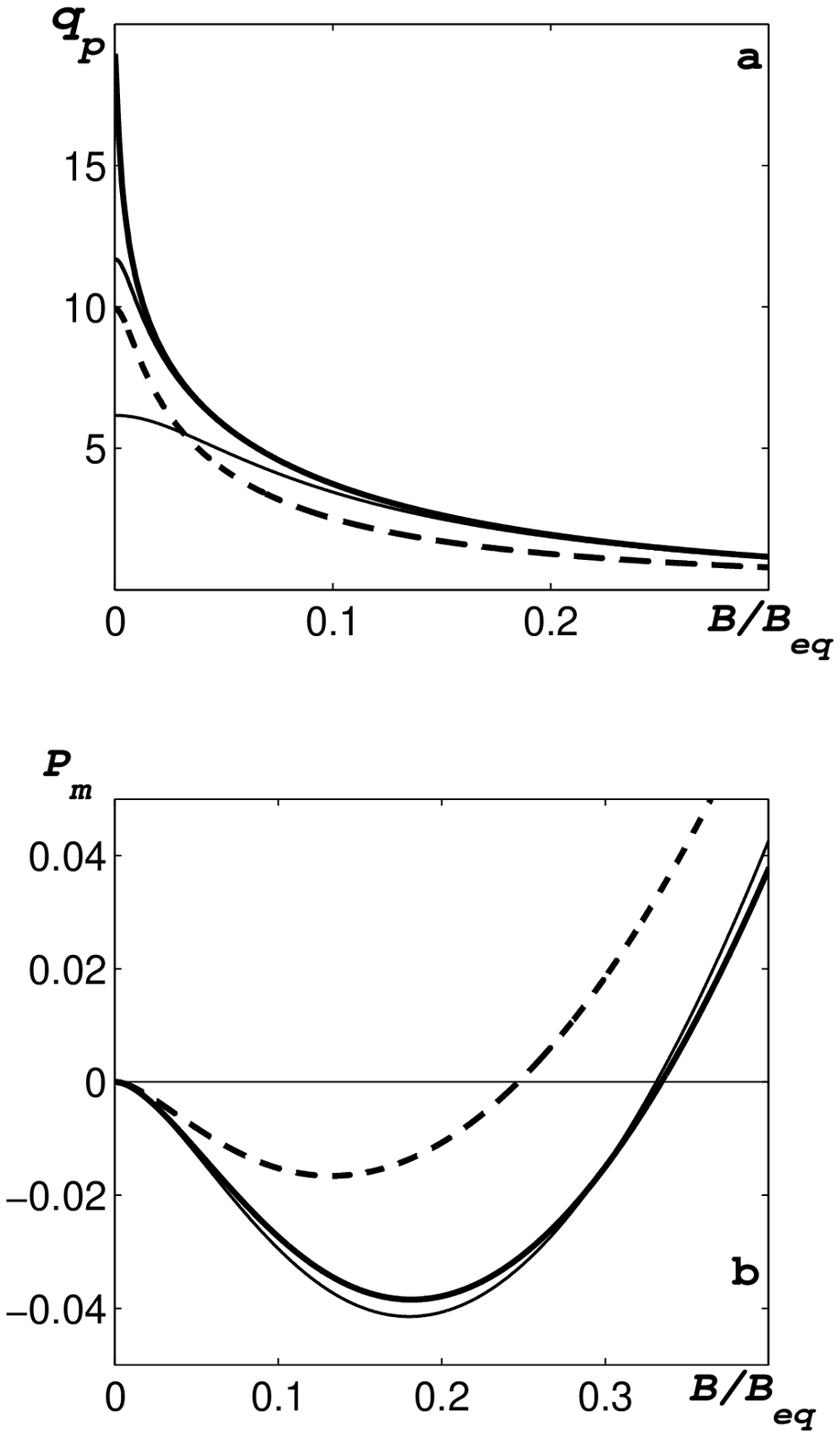}
\caption{\label{Fig3} {\bf (a).} The nonlinear coefficient
$q_p({B})$ for different values of the magnetic Reynolds numbers
${\rm Rm}$: $\, \, {\rm Rm}=10^3$ (thin solid line);  $\, \, {\rm
Rm}=10^6$ (dashed-dotted line); $\, \, {\rm Rm}=10^{10}$ (thick
solid line) for a nonconvective turbulence $(a_\ast = 0)$, and at
${\rm
Rm}=10^6$ (dashed line) for a convective turbulence $(a_\ast = 0.7)$.\\
{\bf (b).} The effective (combined) mean magnetic pressure $P_m({B})
= (1-q_p) {B}^2 / {B}^2_{\rm eq}$ at ${\rm Rm}=10^6$ for a
nonconvective turbulence $(a_\ast = 0)$ (thick solid line), and for
a convective turbulence $(a_\ast = 0.7)$  for the horizontal field
(dashed) and for vertical field (thin solid line).}
\end{figure}

\begin{figure}
\centering
\includegraphics[width=8cm]{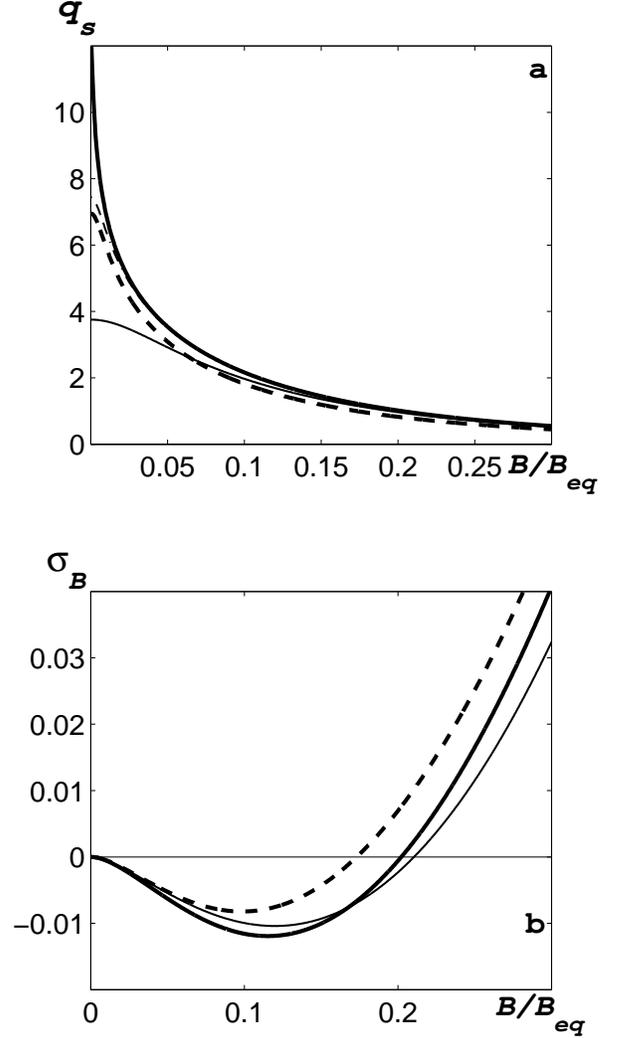}
\caption{\label{Fig4} {\bf (a).} The nonlinear coefficient
$q_s({B})$ for different values of the magnetic Reynolds numbers
${\rm Rm}$: $\, \, {\rm Rm}=10^3$ (thin solid line);  $\, \, {\rm
Rm}=10^6$ (thin dashed-dotted  line); $\, \, {\rm Rm}=10^{10}$
(thick solid) for a nonconvective turbulence $a_\ast = 0$, and at
${\rm Rm}=10^6$ (dashed  line) for a convective turbulence $a_\ast =
0.7$.
\\
{\bf (b).} The effective (combined) mean magnetic tension
$\sigma_{_{B}}({B}) = (1-q_s) {B}^2 / {B}^2_{\rm eq}$ at ${\rm
Rm}=10^6$ for a nonconvective turbulence $(a_\ast = 0)$ (thick solid
line), and for a convective turbulence $(a_\ast = 0.7)$  for the
horizontal field (dashed line) and for vertical field (thin solid).}
\end{figure}

The nonlinear coefficients $q_p(B)$ and $q_s(B)$ in Eq.~(\ref{X3})
for the effective stress tensor are shown in Figs.~3a and~4a for
different values of the magnetic Reynolds numbers. The nonlinear
coefficients $q_p(B)$ and $q_s(B)$ increase with the magnetic
Reynolds numbers in the range of weak mean magnetic fields $(B < 0.1
\, B_{\rm eq})$. On the other hand, the turbulent convection reduces
these nonlinear coefficients in comparison with the case of a
nonconvective turbulence. The asymptotic formulae for the nonlinear
coefficients $q_p(B)$ and $q_s(B)$ are given below. In particular,
for a very weak mean magnetic field, $ B \ll B_{\rm eq} / 4 {\rm
Rm}^{1/4} $, the nonlinear coefficients $q_p(B)$ and $q_s(B)$ are
given by
\begin{eqnarray}
q_p(B) &=& {4 \over 5} \, (1 - \epsilon) \,  \biggl[\ln {\rm Rm} +
{4 \over 45} \biggr] - {8 a_\ast \over 35} (11 - 13 \cos^2 \phi),
\nonumber \\
\label{T40}\\
q_s(B) &=& {8 \over 15} \, (1 - \epsilon) \,  \biggl[\ln {\rm Rm} +
{2 \over 15} \biggr] - {24 a_\ast \over 35} \;,
\label{T41}
\end{eqnarray}
where $\epsilon \equiv \langle {\bf b}^2 \rangle^{(0)} / \langle
{\bf u}^2 \rangle^{(0)}$. For $ { B}_{\rm eq} / 4 {\rm Rm}^{1/4} \ll
B \ll { B}_{\rm eq} / 4 $ these nonlinear coefficients are given by
\begin{eqnarray}
q_p(B) &=& {16 \over 25} \, (1 - \epsilon) \, [5|\ln (4 B)| + 1 + 32
\, B^{2}]
\nonumber \\
&& - {8 a_\ast \over 35} (11 - 13 \cos^2 \phi) \;,
\label{T42}\\
q_s(B) &=& {32 \over 15} \, (1 - \epsilon) \, \biggl[|\ln (4 B)| +
{1 \over 30} + 12  B^{2} \biggr] - {24 a_\ast \over 35} \;,
\nonumber \\
\label{T43}
\end{eqnarray}
and for $  B \gg { B}_{\rm eq} / 4 $ they are given by
\begin{eqnarray}
q_p(B) &=& {1 \over 6  B^2} \, (1 - \epsilon) + {\pi a_\ast \over 80
B} (1 - 5 \cos^2 \phi) \;,
\label{T44}\\
q_s(B) &=& {\pi \over 48  B^3} \, (1 - \epsilon)  + {3 \pi a_\ast
\over 160  B} (1 - 3 \cos^2 \phi) \; .
\label{T45}
\end{eqnarray}
The effective (combined) mean magnetic pressure $P_m(B) = (1-q_p) \,
B^2 / B^2_{\rm eq}$ is shown in Fig.~3b. Inspection of Fig.~3b shows
that the combined mean magnetic pressure $P_m(B) = (1-q_p) \, B^2 /
B^2_{\rm eq}$ vanishes at some value of the mean magnetic field
$B=B_P \sim (0.2 - 0.3) \, B_{\rm eq}$. This causes the following
effect. Let us consider an isolated tube of magnetic field lines.
When $B=B_P$, the combined mean magnetic pressure $P_m(B)=0$, the
fluid pressure and fluid density inside and outside the isolated
tube are the same, and therefore, this isolated tube is in
equilibrium. When $B>B_P$, the combined mean magnetic pressure
$P_m(B)>0$, the fluid pressure and fluid density inside the isolated
tube are smaller than the fluid pressure and fluid density outside
the isolated tube. This results in upwards floating of the isolated
tube. On the other hand, when $B<B_P$, the combined mean magnetic
pressure $P_m(B)<0$, the fluid pressure and fluid density inside the
isolated tube are larger than the fluid pressure and fluid density
outside the isolated tube. Therefore, this isolated magnetic tube
flows down.

The effective (combined) mean magnetic tension $\sigma_{_{B}}(B) =
(1-q_s) B^2 / B^2_{\rm eq}$  is shown in Fig.~4b. The combined mean
magnetic tension $\sigma_{_{B}}(B)$ vanishes at some value of the
mean magnetic field $B=B_S \sim 0.2 \, B_{\rm eq}$ (see Fig.~4b).
When $B>B_S$, the combined mean magnetic tension $\sigma_{_{B}}(B) >
0$, and Alfv\'{e}nic and magneto-sound waves can propagate in the
isolated tubes. On the other hand, when $B<B_S$, the combined mean
magnetic tension $\sigma_{_{B}}(B) < 0$, and Alfv\'{e}nic and
magneto-sound waves cannot propagate in the isolated tubes.

The anisotropic contributions $\sigma_{ij}^A(B)$ to the effective
stress tensor determine the anisotropic mean magnetic tension due to
the turbulent convection. The tensor $\sigma_{ij}^A(B)$ is
characterized by the function $\sigma_{_{A}}(B) = \sigma_{ij}^A(B)
\, e_{ij} = q_e B^2 / B^2_{\rm eq}$. The nonlinear coefficient
$q_e(B)$ and the anisotropic mean magnetic tension,
$\sigma_{_{A}}(B)$ are shown in Fig.~5a and~5b. In the next Section
we show that the anisotropic mean magnetic tension
$\sigma_{_{A}}(B)$ caused by the turbulent convection, strongly
affects the dynamics of the horizontal mean magnetic field.

\begin{figure}
\centering
\includegraphics[width=8cm]{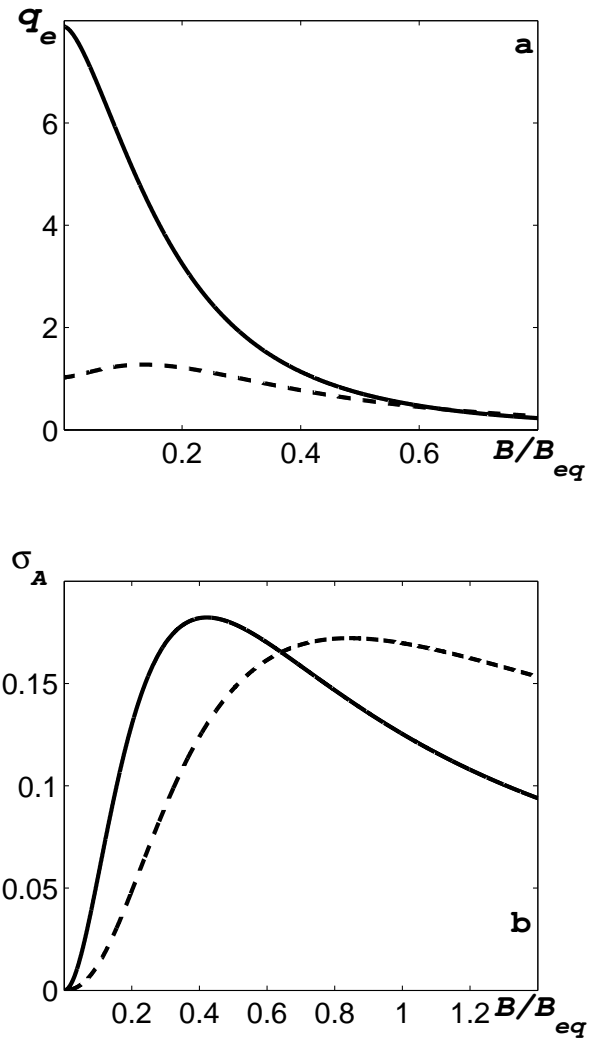}
\caption{\label{Fig5} The nonlinear coefficient $q_e({B})$ versus
the mean magnetic field -- {\bf (a)} and the anisotropic mean
magnetic tension, $\sigma_{_{A}}({B}) = \sigma_{ij}^A e_{ij} = q_e
{B}^2 / {B}^2_{\rm eq}$ -- {\bf (b)} for a convective turbulence
$(a_\ast = 1)$ with the horizontal mean magnetic field (solid line)
and vertical mean magnetic field (dashed line).}
\end{figure}

In this Section we demonstrate that turbulent convection strongly
modifies the large-scale magnetic force. Let us discuss a
possibility for a study of the effect of turbulence on the effective
(combined) mean Lorentz force in the direct numerical simulations.
Consider the mean magnetic field which is directed along the
$x$-axis, i.e., ${\bf B} =  B \, {\bf e}_x$. Let us introduce the
functions $\sigma_x(B)$ and $\sigma_y(B)$:
\begin{eqnarray}
\sigma_x(B) &=&  - {1 \over 2}  \langle {\bf b}^2 \rangle + \langle
b_x^2 \rangle - \rho_0 \langle u_x^2 \rangle \;,
\label{R40}\\
\sigma_y(B) &=&  - {1 \over 2}  \langle {\bf b}^2 \rangle + \langle
b_y^2 \rangle - \rho_0 \langle u_y^2 \rangle \;,
\label{R41}
\end{eqnarray}
which allow us to determine the coefficients $q_p(B)$ and $q_s(B)$
in the effective stress tensor:
\begin{eqnarray}
q_p(B) &=&  {2 \over  B^2} [\sigma_y(B) - \sigma_y(B=0)] \;,
\label{R42}\\
q_s(B)  &=& {1 \over 2} q_p(B) -  {1 \over B^2} [\sigma_x(B) -
\sigma_x(B=0)] \; .
\label{R43}
\end{eqnarray}
Therefore, Eqs.~(\ref{R40})-(\ref{R43}) allows to determine the
effective (combined) mean magnetic force in the direct numerical
simulations.

\section{The large-scale instability}

The modification of the mean magnetic force by the turbulent
convection causes a large-scale instability. In this study we
investigate the large-scale instability of continuous magnetic field
in small-scale turbulent convection and we do not consider buoyancy
of the discrete magnetic flux tubes. In order to study the
large-scale instability in a small-scale turbulent convection we use
the equation of motion (with the effective magnetic force ${\cal
F}_i^{\rm eff} = \nabla_j \sigma_{ij}^{\rm eff}$ determined in
Section IV), the induction equation and the equation for the
evolution of the mean entropy [see Eqs.~(\ref{D5})-(\ref{DD5}) in
Appendix A]. We estimate the growth rate $\gamma$ of this
instability and the frequencies $\omega$ of generated modes
neglecting turbulent dissipative processes  for simplicity's sake.
We also neglect very small Brunt-V\"{a}is\"{a}l\"{a} frequency based
on the gradient of the mean entropy. We seek for the solution of
these equations in the form $\propto \exp(\gamma t + i \omega t - i
{\bf K} \cdot {\bf R}) $.

\subsection{Horizontal mean magnetic field}

First, we study the large-scale instability of a horizontal mean
magnetic field that is perpendicular to the gravity field. Let the
$z$ axis of a Cartesian coordinate system be directed opposite to
the gravitational field, and let the $x$ axis lie along the mean
magnetic field ${\bf B}$. Consider the case $K_x = 0$ that
corresponds to the interchange mode. The dispersive relation for the
instability reads
\begin{eqnarray}
\hat \sigma^2 = \biggl({K_y \, C_{A} \over K \, L_{\rho}} \biggr)^2
\, (Q + i D) \;, \label{DDD7}
\end{eqnarray}
where $\hat \sigma = \gamma + i \omega$, $\, L_\rho$ is the density
stratification length, $L_B$ is the characteristic scale of the mean
magnetic field variations, $C_{A} = B/ \sqrt{\rho_0}$ is the
Alfv\'{e}n speed and
\begin{eqnarray*}
Q &=& \biggl\{1-q_p(y)- y \, q_p'(y) + a_\ast \, \biggl[y
\sigma_{_{A}}''(y) \, \biggl(1 - 2 \, {L_{\rho} \over L_{B}}\biggr)
\nonumber\\
&& + {1 \over 2} \, \sigma_{_{A}}'(y) \, \biggl(1 - 4 \, {L_{\rho}
\over L_{B}}\biggr) \biggr]\biggr\}_{y = B^2}  \,\biggl({L_{\rho}
\over L_{B}} - 1\biggr) \;,
\\
D &=& a_\ast \, K_z \, L_{\rho} \, [\sigma_{_{A}}'(y)]_{y = B^2} \,
\biggl({L_{\rho} \over L_{B}} - 1\biggr) \;,
\end{eqnarray*}
$\sigma_{_{A}}(y) = q_e(y) \, y$, $\, \sigma_{_{A}}'(y) =
d\sigma_{_{A}}(y) / dy$, $\, B$ is measured in the units of the
equipartition field $ B_{\rm eq} = \sqrt{\rho_0} \, u_0 $ and  $K =
\sqrt{K_z^2 + K_y^2}$.

When $Q \geq 0$ the growth rate of perturbations with the frequency
\begin{eqnarray}
\omega = {K_y \over \sqrt{2} \, K} \, \biggl({C_{A} \over L_{\rho}}
\biggr) \, D \, \Big(\sqrt{Q^2 + D^2} + Q \Big)^{- 1/2}\;,
\label{D7}
\end{eqnarray}
is given by
\begin{eqnarray}
\gamma = {K_y \over \sqrt{2} \,K} \, \biggl({C_{A} \over L_{\rho}}
\biggr) \, \Big(\sqrt{Q^2 + D^2} + Q \Big)^{1/2} \; .
\label{D8}
\end{eqnarray}

\begin{figure}
\centering
\includegraphics[width=8cm]{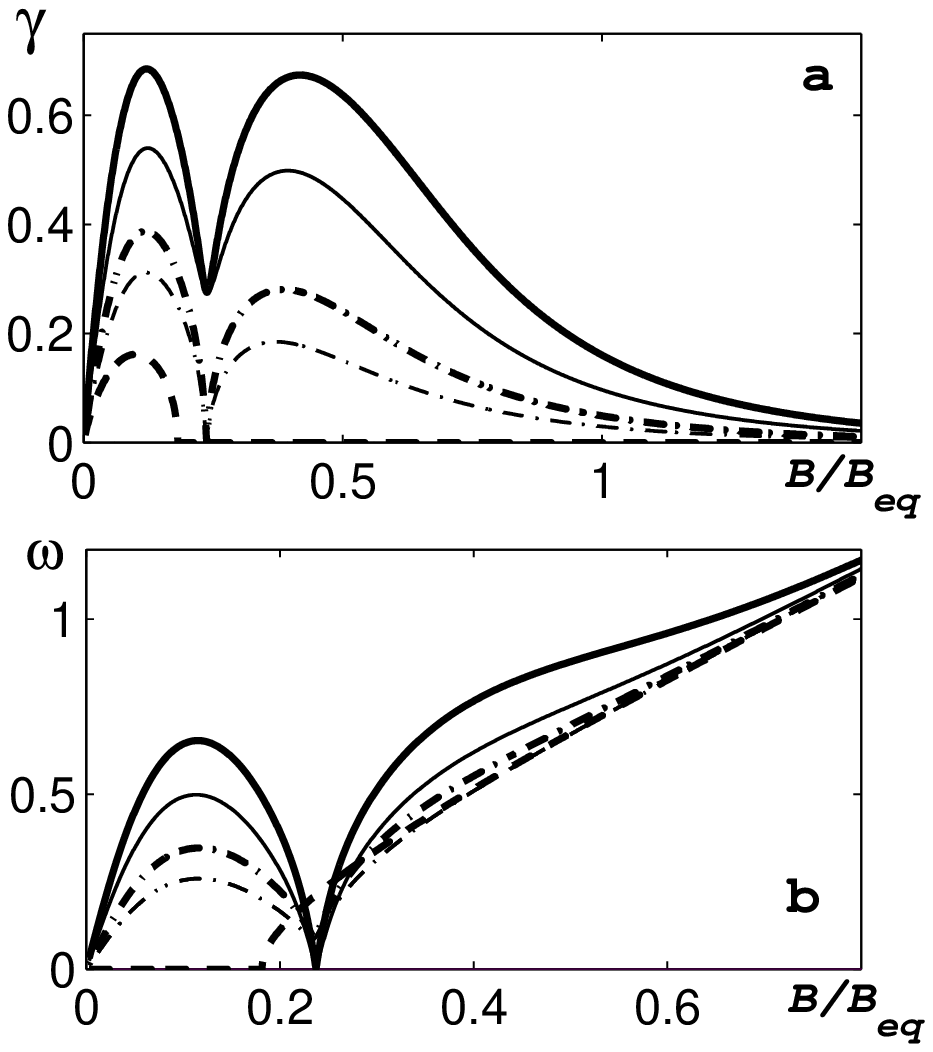}
\caption{\label{Fig6} The growth rate {\bf (a)} of the large-scale
instability and frequency of the generated modes {\bf (b)} for the
horizontal mean magnetic field with $L_{\rho} /L_{B} = 0.2$ versus
$B / B_{\rm eq}$ for different values $K_z$: $\, K_z \, L_{\rho} =
3$ (dashed-dotted line) and $K_z \, L_{\rho} = 5$ (solid line) for
turbulent convection with $a_\ast = 1$ (thick curves)  and $a_\ast =
0.3$ (thin curves), and for a nonconvective turbulence $a_\ast = 0$
(dashed line). Here $\epsilon=0$ and ${\rm Rm}=10^6$.}
\end{figure}

\begin{figure}
\centering
\includegraphics[width=8cm]{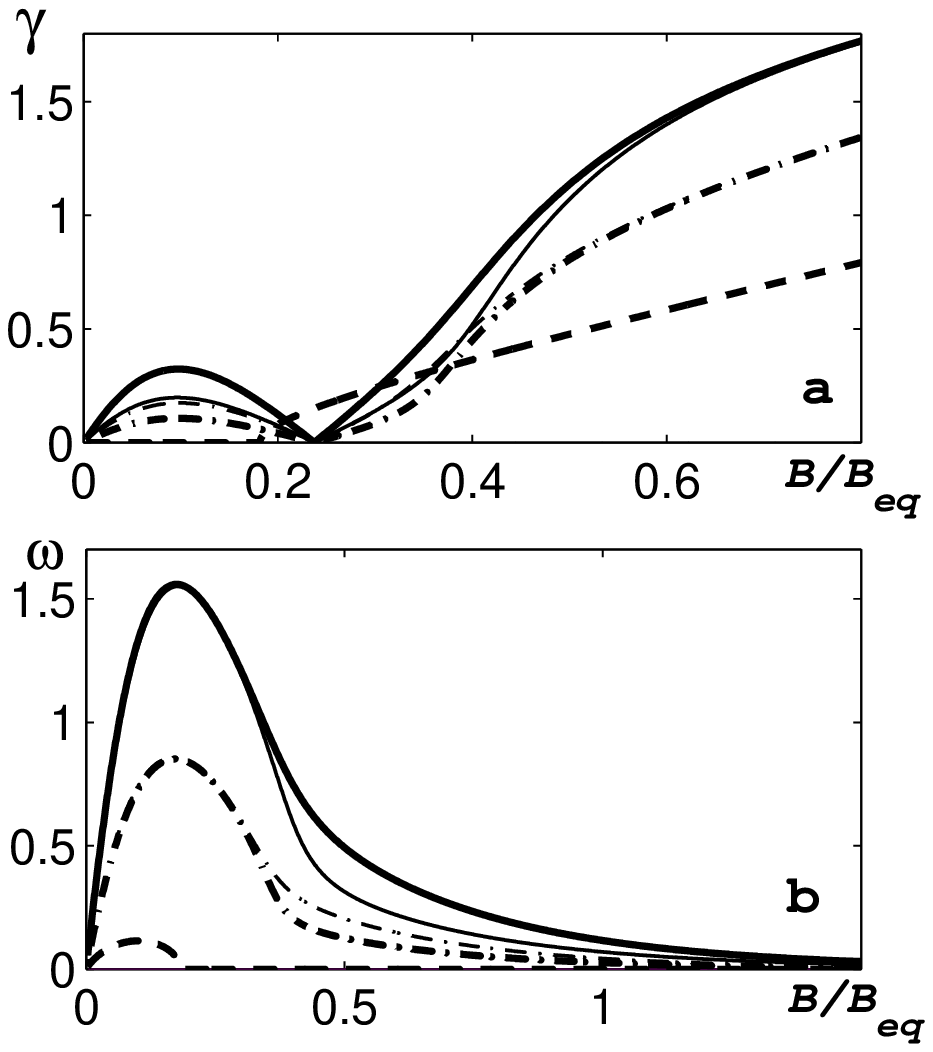}
\caption{\label{Fig7} The growth rate {\bf (a)} of the large-scale
instability and frequency of the generated modes {\bf (b)} for the
horizontal mean magnetic field with $L_{\rho} /L_{B} = 10$ versus $B
/ B_{\rm eq}$ for different values $K_z$: $\, K_z \, L_{\rho} = 3$
(dashed-dotted line) and $K_z \, L_{\rho} = 5$ (solid line) for
turbulent convection with $a_\ast = 1$ (thick curves)  and $a_\ast =
0.3$ (thin curves), and for a nonconvective turbulence $a_\ast = 0$
(dashed line). Here $\epsilon=0$ and ${\rm Rm}=10^6$.}
\end{figure}

When $Q<0$ the growth rate of perturbations with the frequency
\begin{eqnarray}
\omega = - sgn(D) \, {K_y \over \sqrt{2} \,K} \, \biggl({C_{A} \over
L_{\rho}} \biggr) \, \Big(\sqrt{Q^2 + D^2} + |Q| \Big)^{1/2} ,
\label{DQ8}
\end{eqnarray}
is given by
\begin{eqnarray}
\gamma = {K_y \over \sqrt{2} \, K} \, \biggl({C_{A} \over L_{\rho}}
\biggr) \, |D| \, \Big(\sqrt{Q^2 + D^2} + |Q| \Big)^{- 1/2} \; .
\label{DQ7}
\end{eqnarray}
Therefore, in small-scale turbulent convection this large-scale
instability causes excitation of oscillatory modes with growing
amplitude. In a nonconvective turbulence $(a_\ast=0)$ this
large-scale instability is aperiodic.

The growth rate $\gamma$ of the large-scale instability and
frequency $\omega$ of the generated modes for the horizontal mean
magnetic field versus $B / B_{\rm eq}$ for different values $K_z$
and $L_{\rho} /L_{B}$ for a nonconvective and convective turbulence
are shown in Figs.~6 and~7. Here $\gamma$ and $\omega$ are measured
in the units of $t_\ast^{-1}$, where $t_\ast = (L_{\rho} /u_0)
(K/K_y)$.  In the turbulent convection there are two ranges for the
large-scale instability of the horizontal mean magnetic field (when
$Q>0$ and $Q<0)$, while in a nonconvective turbulence $(a_\ast = 0)$
there is only one range for the instability. The first range for the
instability is related to the negative contribution of turbulence to
the effective magnetic pressure for the case of $L_{\rho} < L_{B}$,
while the second range is mainly caused by the anisotropic
contribution $\sigma_{_{A}}(B)$ due to turbulent convection.

In the absence of turbulence (small Reynolds numbers) or turbulent
convection (small Rayleigh numbers) the coefficients $Q=L_{\rho} /
L_{B} - 1$, $\, D=0$, and the criterion for the large-scale
instability, $L_{\rho} > L_{B}$, coincides with that of the Parker's
magnetic buoyancy instability (\cite{P66,P79,G70}). In this case
$\omega=0$, i.e., the oscillatory modes with growing amplitude are
not excited. On the other hand, in a developed turbulent convection
the effective (combined) magnetic pressure becomes negative and the
Parker's magnetic buoyancy instability cannot be excited. However,
the instability due to the modification of the mean magnetic force
by small-scale turbulent convection can be excited even when
$L_{\rho} < L_{B}$, i.e., even in uniform mean magnetic field.

The instability mechanism due to the modification of the mean
magnetic force consists in the following. An isolated tube of
magnetic field lines moving upwards, turns out to be lighter than
the surrounding plasma. This is due to the fact that the decrease of
the magnetic field in the isolated tube caused by its expansion, is
accompanied by an increase of the effective magnetic pressure inside
the tube. Since the effective (combined) magnetic pressure is
negative, this leads a decrease of the density inside the tube. The
arising buoyant force results in the upwards floating of the
isolated tube, i.e. it causes the excitation of the large-scale
instability (see also discussion in \cite{KMR96}).

\subsection{Vertical uniform mean magnetic field}

Consider vertical uniform mean magnetic field, i.e., the magnetic
field is directed along $z$ axis. The growth rate of perturbations
is given by
\begin{eqnarray}
\gamma = C_{A} \, K_z \, \biggl[q_s(y)- 1 - 2y \, q_s'(y) {K_\perp^2
\over K^2} \biggr]^{1/2}_{y = B^2} \;,
\label{D12}
\end{eqnarray}
where ${\bf K}_\perp$ is the  component of the wave vector that is
perpendicular to $z$ axis and $K = \sqrt{K_z^2 + {\bf K}_\perp^2}$.
It follows from Eq.~(\ref{D12}) that the large-scale instability of
the vertical uniform mean magnetic field is caused by the
modification of the mean magnetic tension by small-scale turbulent
convection. When $q_s > 1$ (i.e., when $B < 0.2 B_{\rm eq})$ the
instability occurs for an arbitrary value $K_\perp$, while for $q_s
< 1$ the necessary condition for the instability reads $K_\perp > K
\, \sqrt{\chi}$, where
\begin{eqnarray*}
\chi = \biggl[{1 - q_s(y) \over 2y \, |q_s'(y)|}\biggr]_{y = B^2}
\;,
\end{eqnarray*}
and we take into account that $q_s'(y) < 0$. The growth rate of the
instability is reduced by the turbulent dissipation $\propto
\nu_{_{T}} \, K^2 $. The maximum growth rate $\gamma_{\rm max}$ at a
fixed value of the wave number $K$ (i.e., at a fixed value of the
turbulent dissipation) is attained at $K_z = K_m = K \,
[(1-\chi)/2]^{1/2}$, and it is given by
\begin{eqnarray}
\gamma_{\rm max} = K \, C_A \, (1-\chi) \, \Big[{y \, |q_s'(y)|
\over 2}\Big]^{1/2}_{y = B^2} \;, \label{DD12}
\end{eqnarray}
where $\chi < 1$. The maximum growth rate $\gamma_{\rm max}$ of the
instability for the vertical uniform mean magnetic field versus $B /
B_{\rm eq}$ is plotted in Fig.~8. Here $\gamma_{\rm max}$ is
measured in the units of $u_0 \, K$. The value of $\gamma_{\rm max}$
is larger for a nonconvective turbulence, but the range for the
instability is wider for the turbulent convection (see Fig.~8).

\begin{figure}
\centering
\includegraphics[width=8cm]{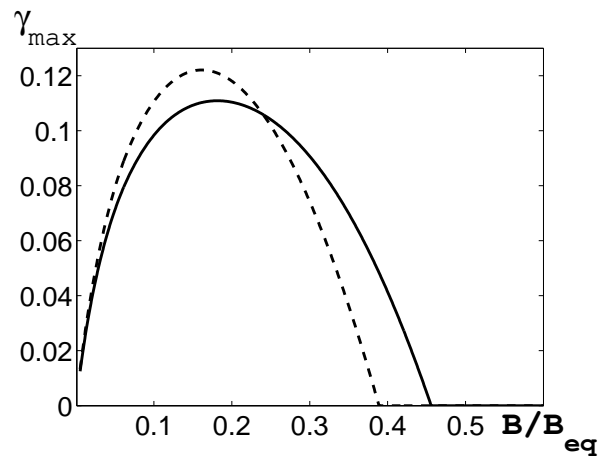}
\caption{\label{Fig8} The maximum growth rate of the large-scale
instability for the vertical mean magnetic field versus $B / B_{\rm
eq}$  for turbulent convection with $a_\ast = 1$ (solid line) and
for a nonconvective turbulence $a_\ast = 0$ (dashed line). Here
$\epsilon=0$ and ${\rm Rm}=10^6$.}
\end{figure}

\section{Discussion}

In the present study we investigate magnetic fluctuations generated
by a tangling of the mean magnetic field in a developed turbulent
convection. When the mean magnetic field $B \ll B_{\rm eq} / 4 {\rm
Rm}^{1/4}$, the energy of magnetic fluctuations depends on magnetic
Reynolds number. We study the modification of the large-scale
magnetic force by turbulent convection. We show that the generation
of magnetic fluctuations in a turbulent convection results in a
decrease of the total turbulent pressure and may cause formation of
the large-scale magnetic structures even in an originally uniform
mean magnetic field. This phenomenon is due to a negative
contribution of the turbulent convection to the effective mean
magnetic force.

The large-scale instability causes the formation of inhomogeneous
magnetic structures. The energy for these processes is supplied by
the small-scale turbulent convection, and this effect can develop
even in an initially uniform magnetic field. In contrast, the
Parker's magnetic buoyancy instability is excited when the density
stratification scale is larger than the characteristic scale of the
mean magnetic field variations (see \cite{P66,P79,G70}). The free
energy in the Parker's magnetic buoyancy instability is drawn from
the gravitational field. The characteristic time of the large-scale
instability is of the order of the Alfv\'{e}n time based on the
large-scale magnetic field.

We study an initial stage of formation of the large-scale magnetic
structures for horizontal and vertical mean magnetic fields relative
to the vertical direction of the gravity field. In the turbulent
convection there are two ranges for the large-scale instability of
the horizontal mean magnetic field. The first range for the
instability is related to the negative contribution of turbulence to
the effective magnetic pressure for the case of $L_{\rho} < L_{B}$,
while the second range for the instability is mainly caused by the
anisotropic contribution of the turbulent convection to the
effective magnetic force. The large-scale instability of the
vertical uniform mean magnetic field is caused by the modification
of the mean magnetic tension by small-scale turbulent convection.
The discussed effects in the present study might be useful for the
understanding of the origin of the sunspot formation.

Since in the present study we neglect very small
Brunt-V\"{a}is\"{a}l\"{a} frequency based on the gradient of the
mean entropy, we do not investigate the large-scale dynamics of the
mean entropy. This problem was addressed in \cite{KM00} whereby the
modification of the mean magnetic force by the turbulent convection
was not taken into account.

In order to study magnetic fluctuations and the modification of the
large-scale Lorentz force by turbulent convection we apply
the spectral $\tau$ approximation (see Sect.~III).
The $\tau$ approach is an universal tool in turbulent transport that allows to obtain closed results and compare them with the results of laboratory experiments, observations and numerical simulations. The $\tau$ approximation reproduces many well-known phenomena found by other methods in turbulent transport of particles and magnetic fields, in turbulent convection and stably stratified turbulent flows (see below).

In turbulent transport, the $\tau$ approximation yields correct formulae for turbulent diffusion, turbulent thermal diffusion and turbulent barodiffusion (see, e.g., \cite{EKR96,BF03}). The phenomenon of turbulent thermal diffusion
(a nondiffusive streaming of particles in the direction of the mean heat flux), has been predicted using the stochastic calculus (the path
integral approach) and the $\tau$ approximation. This phenomenon has been already detected in laboratory experiments in oscillating grids turbulence \cite{EEKR04} and in a multi-fan turbulence generator \cite{EEKR06} in stably and unstably
stratified fluid flows. The experimental results obtained in \cite{EEKR04,EEKR04} are in a good agreement with the theoretical studies performed by means of different approaches (see \cite{EKR96,PM02}).

The $\tau$ approximation reproduces the
well-known $k^{-7/3}$-spectrum of anisotropic velocity fluctuations
in a sheared turbulence (see \cite{EKRZ02}). This spectrum was found previously in analytical, numerical, laboratory studies and was observed in the atmospheric
turbulence (see, e.g., \cite{L67}). In the turbulent boundary layer problems, the
$\tau$-approximation yields correct expressions for turbulent viscosity, turbulent thermal conductivity and the classical heat flux. This approach also describes the counter wind heat flux and the Deardorff's heat flux in convective boundary layers (see \cite{EKRZ02}). These phenomena have been studied previously using different approaches (see, e.g.,  \cite{MY75,Mc90,Z91}).

The theory of turbulent convection \cite{EKRZ02} based on the $\tau$-approximation
explains the recently discovered hysteresis phenomenon in laboratory
turbulent convection \cite{EEKRM06}. The results
obtained using the $\tau$-approximation allow also to explain the most pronounced features of typical semi-organized coherent structures observed in the atmospheric convective boundary layers ("cloud cells" and "cloud streets") \cite{ET85}. The theory \cite{EKRZ02} based on the $\tau$-approximation predicts realistic values of the following parameters: the aspect ratios of structures, the ratios of the
minimum size of the semi-organized structures to the maximum scale
of turbulent motions and  the characteristic lifetime of the
semi-organized structures. The theory \cite{EKRZ02} also predicts excitation of
convective-shear waves propagating perpendicular to the convective
rolls ("cloud streets"). This waves have been observed in the atmospheric
convective boundary layers with cloud streets \cite{ET85}.

A theory \cite{ZEKR07} for stably stratified atmospheric turbulent flows based on the $\tau$-approximation and the budget equations for the key second moments, turbulent kinetic and potential energies and vertical turbulent fluxes of momentum and buoyancy, is in a good agrement with data from atmospheric and laboratory experiments, direct numerical simulations and large-eddy simulations (see detailed comparison in Sect. 5 of \cite{ZEKR07}).

The detailed verification of the $\tau$ approximation in the direct numerical simulations of turbulent transport of passive scalar has been recently
performed in \cite{BK04}. In particular, the results
on turbulent transport of passive scalar obtained using direct
numerical simulations of homogeneous isotropic turbulence have been
compared with that obtained using a closure model based on the
$\tau$ approximation. The numerical and analytical results are in a
good agreement.

In magnetohydrodynamics, the $\tau$ approximation reproduces many
well-known phenomena found by different methods, e.g., the
$\tau$ approximation yields correct formulae
for the $\alpha$-effect \cite{KR80,RK93,RK00,RKR03}, the turbulent
diamagnetic and paramagnetic velocities \cite{Z57,VK83,K91,KR92,RKR03},
the turbulent magnetic diffusion \cite{KR80,VK83,KRP94,RKR03,RK04}, the ${\bf \Omega} {\bf \times} {\bf J}$ effect and the $\kappa$-effect \cite{KR80,RKR03}, the shear-current effect \cite{RK03,RK04,RK07}.

Generation of the large-scale magnetic field in a nonhelical
turbulence with an imposed mean velocity shear has been recently
investigated in \cite{BH05} using direct numerical simulations. The results of these
numerical simulations are in a good agreement with the theoretical
predictions based on the $\tau$ approximation (see \cite{RK03,RK04,RK07})
and with the numerical solutions of the nonlinear dynamo equations performed in
\cite{BS05B,RKL06} (see detailed comparison in \cite{RK07}).

The validity of the $\tau$ approximation has been tested in the context of dynamo theory, in direct numerical simulations in \cite{BSM05}. The alpha effect
in mean field dynamo theory becomes proportional to a relaxation time scale multiplied by the difference between kinetic and
current helicities. It is shown in \cite{BSM05} that the value of the relaxation time is positive and, in units of the turnover time at the forcing
wavenumber, it is of the order of unity. Kinetic and current helicities are shown in \cite{BSM05} to be dominated by large scale properties of the flow.
Recent studies in \cite{SSB07} of the nonlinear alpha effect showed that in the
limit of small magnetic and hydrodynamic Reynolds numbers, both the second order correlation approximation (or first-order smoothing approximation) and the $\tau$ approximation give identical results. This is also supported by simulations \cite{BS07} of isotropically forced helical turbulence whereby the contributions to kinetic and magnetic alpha effects are computed. The study performed in \cite{BS07} provides an extra piece of evidence that the $\tau$ approximation is a
useable formalism for describing simulation data and for
predicting the behavior in situations that are not yet accessible
to direct numerical simulations.

\begin{acknowledgments}
We have benefited from stimulating discussions with Axel Brandenburg
and Alexander Schekochihin. This work has been initiated  during our
visit to the Isaac Newton Institute for Mathematical Sciences
(Cambridge) in the framework of the programme "Magnetohydrodynamics
of Stellar Interiors".
\end{acknowledgments}

\appendix

\section{Velocity and magnetic fluctuations in turbulent convection}

In order to study the velocity and magnetic fluctuations with a
nonzero mean magnetic field and to derive the effective stress
tensor in the turbulent convection, we use a mean field approach in
which the magnetic and velocity fields, and entropy are decomposed
into the mean and fluctuating parts, where the fluctuating parts
have zero mean values. The equations for fluctuations of the fluid
velocity, entropy and the magnetic field are given by
\begin{eqnarray}
{1 \over \sqrt{\rho_0}} {\partial {\bf v}({\bf x},t) \over
\partial t} &=& - \bec{\nabla} \biggl({p \over
\rho_0} \biggr) - {{\bf g} \over \sqrt{\rho_0}} \, s  + {1 \over
\sqrt{\rho_0}} \biggl[({\bf b} \cdot \bec{\nabla}) {\bf H}
\nonumber \\
&& + ({\bf H} \cdot \bec{\nabla}){\bf b} + \, {\Lambda_\rho \over 2}
[2{\bf e}({\bf b} \cdot {\bf H})
\nonumber \\
&&  - ({\bf b} \cdot {\bf e}) {\bf H}]
\biggr] + {\bf v}^N \,,
\label{B1} \\
{\partial {\bf b}({\bf x},t) \over \partial t} &=& ({\bf H} \cdot
\bec{\nabla}){\bf v} - ({\bf v} \cdot \bec{\nabla}) {\bf H} +
{\Lambda_\rho \over 2} [{\bf v} ({\bf H} \cdot {\bf e})
\nonumber \\
&& - \, {\bf H} ({\bf v} \cdot {\bf e})] + {\bf b}^N \,,
\label{B2} \\
{\partial s({\bf x},t) \over \partial t} &=& - {\Omega_{b}^{2} \over
g} ({\bf v} \cdot {\bf e}) + s^N \,, \label{B3}
\end{eqnarray}
where we used new variables $({\bf v}, \, s, \, {\bf H})$ for
fluctuating fields ${\bf v} = \sqrt{\rho_0} \, {\bf u} $ and $s =
\sqrt{\rho_0} \, s'$, and also for the mean field ${\bf H} = {\bf B}
/ \sqrt{\rho_0}$. Here ${\bf B}$ is the mean magnetic field,
$\rho_0$ is the fluid density, ${\bf e}$ is the vertical unit vector
directed opposite to the gravity field, $\Omega_{b}^{2} = - {\bf g}
\cdot \bec{\nabla} S$ is the Brunt-V\"{a}is\"{a}l\"{a} frequency,
$S$ is the mean entropy, ${\bf g}$ is the acceleration of gravity,
${\bf u}$, $\, {\bf b}$ and $s'$ are fluctuations of velocity,
magnetic field and entropy (we have not used new variables for
magnetic fluctuations), ${\bf v}^{N}$, $\, {\bf b}^{N}$ and $\,
s^{N}$ are the nonlinear terms which include the molecular viscous
and diffusion terms, $p = p' + \sqrt{\rho_0} \, ({\bf H} \cdot {\bf
b})$ are the fluctuations of total pressure, $p'$ are the
fluctuations of fluid pressure.

Equations (\ref{B1})-(\ref{B3}) for fluctuations of fluid velocity,
entropy and magnetic field are written in the anelastic
approximation, which is a combination of the Boussinesq
approximation and the condition ${\rm div} \, (\rho_0 \, {\bf u}) =
0$. The equation, ${\rm div} \, {\bf u} = \Lambda_\rho ({\bf u}
\cdot {\bf e})$, in the new variables reads: ${\rm div} \, {\bf v} =
(\Lambda_\rho / 2) ({\bf v} \cdot {\bf e})$, where $\bec{\nabla}
\rho_0 / \rho_0 = - \Lambda_\rho {\bf e}$. The quantities with the
subscript $ "0" $ correspond to the hydrostatic nearly isentropic
basic reference state, i.e., $\bec{\nabla} P_{0} = \rho_{0} \, {\bf
g} $ and $ {\bf g} \cdot [(\tilde \gamma P_{0})^{-1} \, \bec{\nabla}
P_{0} - \rho_0^{-1} \bec{\nabla} \rho_0] \approx 0$, where $\tilde
\gamma$ is the specific heats ratio and $P_{0}$ is the fluid
pressure in the basic reference state. The turbulent convection is
regarded as a small deviation from a well-mixed adiabatic reference
state.

Using Eqs.~(\ref{B1})-(\ref{B3}) and performing the procedure
described in Section III we derive equations for the two-point
second-order correlation functions of the velocity fluctuations
$f_{ij} = \langle v_i \, v_j\rangle$, the magnetic fluctuations
$h_{ij} = \langle b_i \, b_j \rangle$, the entropy fluctuations
$\Theta = \langle s \, s\rangle$, the cross-helicity $g_{ij} =
\langle b_i \, v_j \rangle$, the turbulent heat flux $F_{i} =
\langle s \, v_i \rangle$ and $G_{i} = \langle s \, b_i \rangle$.
The equations for these correlation functions are given by
\begin{eqnarray}
{\partial f_{ij}({\bf k}) \over \partial t} &=& {\rm i}\,({\bf k}
{\bf \cdot} {\bf H}) \Phi_{ij} + \hat{\cal N} f_{ij} \;,
\label{B6} \\
{\partial h_{ij}({\bf k}) \over \partial t} &=& - {\rm i}\,({\bf
k}{\bf \cdot} {\bf H}) \Phi_{ij} + \hat{\cal N}h_{ij} \;,
\label{B7} \\
{\partial g_{ij}({\bf k }) \over \partial t} &=& {\rm i}\,({\bf k}
{\bf \cdot} {\bf H}) [f_{ij}({\bf k}) - h_{ij}({\bf k})]
\nonumber \\
&& + g e_n P_{jn}(k) G_{i}(-{\bf k}) + \hat{\cal N} g_{ij} \;,
\label{B8} \\
{\partial F_{i}({\bf k}) \over \partial t} &=& - {\rm i}\,({\bf k}
{\bf \cdot} {\bf H}) G_{i}({\bf k}) + g e_n P_{in}(k) \Theta({\bf
k}) + \hat{\cal N} F_{i} \;,
\nonumber \\
\label{B9} \\
{\partial G_{i}({\bf k}) \over \partial t} &=& - {\rm i}\,({\bf k}
{\bf \cdot} {\bf H}) F_{i}({\bf k}) + \hat{\cal N} G_{i} \;,
\label{B10} \\
{\partial \Theta({\bf k}) \over \partial t} &=& - {\Omega_b^2 \over
g} F_{z}({\bf k}) + \hat{\cal N} \Theta \;,
\label{B11}
\end{eqnarray}
(see for details \cite{RK06}), where $\Phi_{ij}({\bf k}) =
g_{ij}({\bf k}) - g_{ji}(-{\bf k})$, $\, \hat{\cal N} f_{ij} = g e_n
[P_{in}(k) F_{j}({\bf k}) + P_{jn}(k) F_{i}(-{\bf k})] + \hat{\cal
N} \tilde f_{ij}$, and $ \hat{\cal N}\tilde f_{ij}$, $\, \hat{\cal
N}h_{ij}$, $\, \hat{\cal N}g_{ij}$, $\, \hat{\cal N}F_{i}$, $\,
\hat{\cal N}G_{i}$ and $\hat{\cal N}\Theta$ are the third-order
moment terms appearing due to the nonlinear terms. The terms $\sim
F_i$ in the tensor $ \hat{\cal N}f_{ij}$ can be considered as a
stirring force for the turbulent convection. Note that a stirring
force in the Navier-Stokes turbulence is an external parameter.

We split the tensor of magnetic fluctuations into nonhelical,
$h_{ij}$, and helical, $h_{ij}^{(H)},$ parts. The helical part
$h_{ij}^{(H)}$ depends on the magnetic helicity, and it is
determined by the dynamic equation which follows from the magnetic
helicity conservation arguments. We also split all second-order
correlation functions into symmetric and antisymmetric parts with
respect to the wave vector ${\bf k}$, e.g., $f_{ij} = f_{ij}^{(s)} +
f_{ij}^{(a)}$, where the tensors $f_{ij}^{(s)} = [f_{ij}({\bf k}) +
f_{ij}(-{\bf k})] / 2$ describes the symmetric part of the tensor
and $f_{ij}^{(a)} = [f_{ij}({\bf k}) - f_{ij}(-{\bf k})] / 2$
determines the antisymmetric part of the tensor. We use the spectral
$\tau$ approximation (see Eq.~(\ref{AAC3}) in Sect.~III). We assume
also that the characteristic time of variation of the mean magnetic
field ${\bf B}$ is substantially larger than the correlation time
$\tau(k)$ for all turbulence scales. This allows us to get a
stationary solution for the equations for the second-order moments:
\begin{eqnarray}
f_{ij}^{(s)}({\bf k}) &\approx& {1 \over 1 + 2 \psi} [(1 + \psi)
f_{ij}^{(0s)}({\bf k}) + \psi h_{ij}^{(0s)}({\bf k})
\nonumber \\
&& - 2 \psi \tau g e_n P_{in}(k) F_{j}^{(s)}({\bf k})]  \;,
\label{B22}\\
h_{ij}^{(s)}({\bf k}) &\approx& {1 \over 1 + 2 \psi} [\psi
f_{ij}^{(0s)}({\bf k}) + (1 + \psi) h_{ij}^{(0s)}({\bf k})
\nonumber \\
&& + \psi \tau g e_n P_{in}(k) F_{j}^{(s)}({\bf k})] \;,
\label{B24}\\
g_{ij}^{(a)}({\bf k}) &\approx& {i \tau ({\bf k} {\bf \cdot} {\bf
H}) \over 1 + 2 \psi} [f_{ij}^{(0s)}({\bf k}) - h_{ij}^{(0s)}({\bf
k})
\nonumber \\
&& + \tau g e_n P_{in}(k) F_{j}^{(s)}({\bf k})]  \; .
\label{B26}\\
F_{i}^{(s)}({\bf k}) &\approx& {F_{i}^{(0s)}({\bf k}) \over 1 +
\psi/2} \;,
\label{B20B}\\
G_{i}^{(a)}({\bf k}) &\approx& - i \tau ({\bf k} {\bf \cdot} {\bf
H}) F_{i}^{(s)}({\bf k}) \;, \label{B21B}
\end{eqnarray}
(see for details \cite{RK06}), where $ \psi({\bf k}) = 2 (\tau \,
{\bf k} {\bf \cdot} {\bf H})^2 $ and we neglected terms $\sim
O(\Omega_b^2)$. In Eqs. (\ref{B22})-(\ref{B21B}) we neglected also
the large-scale spatial derivatives. The correlation functions $
f_{ij}^{(a)}$, $\, h_{ij}^{(a)}$,$\, g_{ij}^{(s)}$ $\, F_{i}^{(a)}$
and $\, G_{i}^{(s)}$ vanish because they are proportional to the
first-order spatial derivatives. Equations~(\ref{B22}) and
(\ref{B24}) yield
\begin{eqnarray}
f_{ij}^{(s)}({\bf k}) &+& h_{ij}^{(s)}({\bf k}) = f_{ij}^{(0s)}({\bf
k}) + h_{ij}^{(0s)}({\bf k})
\nonumber \\
&&- {\psi \over 1 + 2 \psi} \tau g e_n P_{in}(k) F_{j}^{(s)}({\bf
k}) \; . \label{BB27}
\end{eqnarray}
Therefore, when the mean heat flux $F_{i}^{(0s)}$ in the background
turbulence is zero (i.e., for the nonconvective turbulence), we
obtain
\begin{eqnarray}
f_{ij}^{(s)}({\bf k}) + h_{ij}^{(s)}({\bf k}) = f_{ij}^{(0s)}({\bf
k}) + h_{ij}^{(0s)}({\bf k}) \; . \label{B27}
\end{eqnarray}
This is in agreement with the fact that a uniform mean magnetic
field performs no work on the turbulence (without mean heat flux).
It can only redistribute the energy between hydrodynamic
fluctuations and magnetic fluctuation. A change of the total energy
of fluctuations is caused by a nonuniform mean magnetic field. For
the integration in $ {\bf k} $-space of these second moments we have
to specify a model for the background turbulent convection (with
zero mean magnetic field, $ {\bf B} = 0)$. Here we use the following
model of the background turbulent convection [denoted with the
superscript $(0)$]:
\begin{eqnarray}
f_{ij}^{(0)}({\bf k}) &=& \rho_0 \, \langle {\bf u}^2 \rangle^{(0)}
\, W(k) \, P_{ij}({\bf k}) \;,
\label{K1} \\
h_{ij}^{(0)}({\bf k}) &=& \langle {\bf b}^2 \rangle^{(0)} \,  W(k)
\, P_{ij}({\bf k}) \;,
\label{KK1} \\
F^{(0)}_{i}({\bf k}) &=& 3 \rho_0 \, \langle u_i \, s' \rangle^{(0)}
\, W(k) \, e_j \, P_{ij}({\bf k}) \;,
\label{K2} \\
\Theta^{(0)}({\bf k}) &=&  2 \rho_0 \, \langle s'^2 \rangle^{(0)} \,
W(k) \;, \label{K3}
\end{eqnarray}
where $P_{ij}({\bf k}) = \delta_{ij} - k_{i} k_{j} / k^{2}$, $\,
W(k) = E(k) / 8 \pi k^{2}$, $\, \tau(k) = 2 \tau_{0} \bar \tau(k)$,
$\, E(k) = - d \bar \tau(k) / dk$, $\, \bar \tau(k) = (k /
k_{0})^{1-q}$, $\, 1 < q < 3$ is the exponent of the kinetic energy
spectrum (e.g., $q = 5/3$ for Kolmogorov spectrum), $k_{0} = 1 /
l_{0}$ and $\tau_{0} = l_{0} / u_{0} $. Note also that
$g_{ij}^{(0)}({\bf k}) = 0$ and $G_{i}^{(0)}({\bf k}) = 0$.

This procedure allows to study magnetic fluctuations with a nonzero
mean magnetic field and to derive the effective stress tensor in the
turbulent convection (see Section III). In particular, integration
in ${\bf k}$ space in Eq.~(\ref{B24}) yields the energy of magnetic
fluctuations
\begin{eqnarray}
\langle {\bf b}^2 \rangle &=& \langle {\bf b}^2 \rangle^{(0)} + {1
\over 12} (\langle {\bf v}^2 \rangle^{(0)} - \langle {\bf b}^2
\rangle^{(0)}) \, \biggl[6 - 3 A_{1}^{(0)}(4{B})
\nonumber \\
&& - A_{2}^{(0)}(4{B})] \biggr] + {a_\ast \over 6} \langle {\bf v}^2
\rangle^{(0)} \Big[2 \Psi \{A_{1}\}
\nonumber \\
&& + (1 + 3 \cos^2 \phi) \Psi \{A_{2}\} \Big]  \;, \label{D1}
\end{eqnarray}
where $\Psi \{X\} = X^{(1)}(2  B) - X^{(1)}(4  B)$, $\, \phi$ is the
angle between the vertical unit vector ${\bf e}$ and the mean
magnetic field ${\bf B}$, the functions $A_{n}^{(0)}(y)$ are given
by Eqs.~(\ref{XX27}), (\ref{X40})-(\ref{X41}) and the functions
$A_{n}^{(1)}(y)$ are given by Eq.~(\ref{X27}). The magnetic stress
tensor is given by Eqs.~(\ref{X3}), where the anisotropic
contribution $\sigma_{ij}^A$ to the magnetic stress tensor is
determined by
\begin{eqnarray}
\sigma_{ij}^A &=& {1 \over 2} \biggl[ e_{ij} \Psi \{2 C_{1}^{(1)} +
A_{1}^{(1)} + A_{2}^{(1)}\} +  \cos \phi \, (e_i \beta_j
\nonumber \\
&& + e_j \beta_i) \, \Psi \{2 C_{3}^{(1)} - A_{2}^{(1)}\} \biggr]
\;, \label{X4}
\end{eqnarray}
and the nonlinear coefficients $q_p(B)$ and $q_s(B)$ are given by
\begin{eqnarray}
q_p(B) &=& {1 \over 12  B^2} \, \biggl[(1 - \epsilon) \,
[A_{1}^{(0)}(0) - A_{1}^{(0)}(4  B)
\nonumber \\
&& - A_{2}^{(0)}(4  B)] + 2 a_\ast [ \Psi \{6 C_{1}^{(1)} - 2
A_{1}^{(1)} - A_{2}^{(1)}\}
\nonumber \\
&& + \cos^2 \phi \, \Psi \{6 C_{3}^{(1)} + A_{2}^{(1)} \} ] \biggr]
\;,
\label{X1}\\
q_s(B) &=& - {1 \over 12  B^2} \, \biggl[(1 - \epsilon) \,
A_{2}^{(0)}(4  B)   + 6 a_\ast [\Psi \{C_{3}^{(1)}\}
\nonumber \\
&& + \cos^2 \phi \, \Psi \{C_{2}^{(1)}\} ] \biggr] \; .
\label{X2}
\end{eqnarray}
The asymptotic formulas for these coefficients are given by
Eqs.~(\ref{T40})-(\ref{T45}).

In order to study the large-scale instability we use the equation of
motion, the induction equation and the equation for the mean
entropy:
\begin{eqnarray}
&& {D {U}_i \over D t} = - \nabla_i \biggl({\tilde P_{\rm tot} \over
\rho_0} \biggr) + {1 \over \rho_0} \biggl[ (1-q_p(B)) {{\bf B}^2
\over 2} \Lambda_\rho e_i
\nonumber\\
&& + ({\bf B} \cdot {\bf \nabla}) [(1-q_s(B)) {B}_i] + \nabla_j [2
\rho_0 \nu_{_{T}}(B) \, (\partial  U)_{ij}]
\nonumber\\
&& + \nabla_j \sigma_{ij}^A  \biggr] - {\bf g} \, S - \nu_{_{T}}(B)
\, \Lambda_\rho e_i \, {\rm div} \, {\bf U}  \;,
\label{D5}\\
&& {\partial {\bf B}\over \partial t} = {\bf \nabla} {\bf \times}
[{\bf U} \times{\bf B} - \eta_{_{T}}(B) \, ({\bf \nabla} {\bf
\times} {\bf B})] \;,
\label{DDD5}\\
&& {D S \over D t} =- {\bf \nabla} {\bf \cdot} \hat {\bf F}^{(s)}
\;, \label{DD5}
\end{eqnarray}
where $\eta_{_{T}}(B)$ is the turbulent magnetic diffusion, $ D / D
t =\partial / \partial t + ({\bf U} {\bf \cdot} {\bf \nabla})$, and
$\hat {\bf F}^{(s)} = - \kappa_{ij}^{(T)}(B) {\bf \nabla} S$ is the
turbulent heat flux, $\kappa_{ij}^{(T)}$ is the tensor for the
nonlinear turbulent thermal diffusivity, and
\begin{eqnarray*}
\tilde P_{\rm tot} &=& P_k + (1-q_p(B)) \, {{\bf B}^2 \over 2} -
\nu_{_{T}}(B) \, \rho_0 \, {\rm div} \,  {\bf U} \;,
\end{eqnarray*}
$P_k$ is the mean fluid pressure, $2\, (\partial  U)_{ij} = \nabla_i
\, U_j + \nabla_j \, U_i$ and $\nu_{_{T}}(B)$ is the turbulent
viscosity. The turbulent viscosity $\nu_{_{T}}(B)$ in Eq.~(\ref{D5})
is given by
\begin{eqnarray}
\nu_{_{T}}(B) &=& \nu_{_{T}} \biggl[2 A_1^{(1)} + (1 + \epsilon)
A_2^{(1)} - 2 (1 - \epsilon) H\{A_1\}
\nonumber\\
&& - {1 \over 6} \biggl( (1 - 29 \epsilon) C_1^{(1)} - 4 (7 - 8
\epsilon) H\{C_1\}
\nonumber\\
&& + (1 - 3 \epsilon) G\{C_1\} - 2 (1 - \epsilon) Q\{C_1\} \biggr)
\biggr]_{y=4  B} \; .
\nonumber\\
\label{D16}
\end{eqnarray}
The explicit form of the functions $H\{X\}$, $\, G\{X\}$ and
$Q\{X\}$ is given in \cite{RK04}. The asymptotic formula for
$\nu_{_{T}}(B)$ for a weak mean magnetic field, ${B} \ll {B}_{\rm
eq} / 4$, is given by $\nu_{_{T}}(B) = \nu_{_{T}} (1 + 2 \epsilon)$,
and for ${B} \gg {B}_{\rm eq} / 4$ it is given by $\nu_{_{T}}(B) =
(\nu_{_{T}} / 4 B) \, (1 + \epsilon)$. The turbulent heat flux $\hat
F_{i}^{(s)}({\bf B}) = - \kappa_{ij}^{(T)}({\bf B}) \nabla_j S $ and
the tensor for the nonlinear turbulent thermal diffusivity:
\begin{eqnarray}
\kappa_{ij}^{(T)}({\bf B}) &=& {\kappa_{\ast}^{(T)} \over 4}
[(2A_{1}^{(0)}(2  B) + A_{2}^{(0)}(2  B)) \, \delta_{ij}
\nonumber \\
&&- A_{2}^{(0)}(2  B) \, \beta_{ij}] \;, \label{S56}
\end{eqnarray}
where $\kappa_{\ast}^{(T)} = u_0 l_0 / 3$, $\, \beta_{ij} = B_i B_j
/  B^2 $.  The asymptotic formula for $\kappa_{ij}^{(T)}({\bf B})$
for $ B \ll { B}_{\rm eq} / 2 {\rm Rm}^{1/4}$ reads
\begin{eqnarray}
\kappa_{ij}^{(T)}({\bf B}) = {\kappa_{\ast}^{(T)} \over 20} [2 (10 -
\beta^2 \ln {\rm Rm})  \, \delta_{ij} + \beta^2 \ln {\rm Rm} \,
\beta_{ij}] \;,
\nonumber \\
\label{S57}
\end{eqnarray}
where $\beta = \sqrt{8}  B / {B}_{\rm eq}$. When ${B}_{\rm eq} / 2
{\rm Rm}^{1/4} \ll  B \ll {B}_{\rm eq} / 2$, the function
$\kappa_{ij}^{(T)}({\bf B})$ is
\begin{eqnarray}
\kappa_{ij}^{(T)}({\bf B}) = {\kappa_{\ast}^{(T)} \over 5} [(5 - 2
\beta^2 |\ln \beta|)  \, \delta_{ij} + \beta^2 |\ln \beta| \,
\beta_{ij}] \;,
\nonumber \\
\label{S58}
\end{eqnarray}
and when $ B \gg {B}_{\rm eq} / 2$, it is
\begin{eqnarray}
\kappa_{ij}^{(T)}({\bf B}) = \kappa_{\ast}^{(T)} {\sqrt{2} \pi \over
4 \beta} (\delta_{ij} + \beta_{ij}) \; .
\label{S59}
\end{eqnarray}
In order to integrate in Eqs. (\ref{B22})-(\ref{B21B}) over the
angles in $ {\bf k} $-space we used the following identity:
\begin{eqnarray*}
\bar K_{ij} &=& \int {k_{ij} \sin \theta \over 1 + a \cos^{2}
\theta} \,d\theta \,d\varphi = \bar A_{1} \delta_{ij} + \bar A_{2}
\beta_{ij} \;,
\\
\bar K_{ijmn} &=& \int {k_{ijmn} \sin \theta \over 1 + a \cos^{2}
\theta} \,d\theta \,d\varphi
\nonumber \\
&=& \bar C_{1} (\delta_{ij} \delta_{mn} + \delta_{im} \delta_{jn} +
\delta_{in} \delta_{jm}) + \bar C_{2} \beta_{ijmn}
\nonumber \\
&& + \bar C_{3} (\delta_{ij} \beta_{mn} + \delta_{im} \beta_{jn} +
\delta_{in} \beta_{jm} + \delta_{jm} \beta_{in}
\nonumber \\
&&+ \delta_{jn} \beta_{im} + \delta_{mn} \beta_{ij}) \;,
\\
\bar H_{ijmn}(a) &=& \int {k_{ijmn} \sin \theta \over (1 + a
\cos^{2} \theta)^{2} } \,d\theta \,d\varphi
\\
&=& \bar K_{ijmn}(a) + a {\partial \over \partial a} \bar
K_{ijmn}(a) \;,
\\
\bar G_{ijmn}(a) &=& \int {k_{ijmn} \sin \theta \over (1 + a
\cos^{2} \theta)^{3} } \,d\theta \,d\varphi
\\
&=& \bar H_{ijmn}(a) + {a \over 2} {\partial \over
\partial a} \bar H_{ijmn}(a) \;,
\end{eqnarray*}
where $ a = \beta^2 / \bar \tau(k) ,$ and
\begin{eqnarray*}
\bar A_{1} &=& {2 \pi \over a} \biggl[(a + 1) {\arctan (\sqrt{a})
\over \sqrt{a}} - 1 \biggr] \;,
\\
\bar A_{2} &=& - {2 \pi \over a} \biggl[(a + 3) {\arctan (\sqrt{a})
\over \sqrt{a}} - 3 \biggr] \;
\\
\bar C_{1} &=& {\pi \over 2a^{2}} \biggl[(a + 1)^{2} {\arctan
(\sqrt{a}) \over \sqrt{a}} - {5 a \over 3} - 1 \biggr]  \;,
\\
\bar C_{2} &=& \bar A_{2} - 7 \bar A_{1} + 35 \bar C_{1} \;,
\\
\bar C_{3} &=& \bar A_{1} - 5 \bar C_{1} \; .
\end{eqnarray*}
(for details, see \cite{RK04}). The functions $A_{n}^{(m)}(\beta)$
are given by
\begin{eqnarray}
A_{n}^{(0)}(\beta) &=& {3 \beta^{2} \over \pi} \int_{\beta}^{\beta
{\rm Rm}^{1/4}} {\bar A_{n}(X^{2}) \over X^{3}} \,d X  \;,
\label{XX27}\\
A_{n}^{(1)}(\beta) &=& {3 \beta^{4} \over \pi} \int_{\beta}^{\beta
{\rm Rm}^{1/4}} {\bar A_{n}(X^{2}) \over X^{5}} \,d X  \;,
\label{X27}
\end{eqnarray}
and similarly for $C_{n}^{(m)}(\beta)$, where $ X^{2} = \beta^{2} (k
/ k_{0})^{2/3} = a$. The explicit form of the functions $
A_{n}^{(m)}(\beta) $ and $ C_{n}^{(m)}(\beta) $ for $m = 1; \, 2$
are given in \cite{RK04}, and the functions $A_{1}^{(0)}(\beta)$ and
$A_{2}^{(0)}(\beta)$ are given by
\begin{eqnarray}
A_{1}^{(0)}(\beta) &=& {1 \over 5} \biggl[2 + 2 {\arctan \beta
\over \beta^3} (3 + 5 \beta^{2}) - {6 \over \beta^{2}}  -
\beta^{2} \ln {\rm Rm}
\nonumber \\
& & - 2 \beta^{2} \ln \biggl({1 + \beta^{2} \over 1 + \beta^{2}
\sqrt{\rm Rm}}\biggr) \biggr] \;,
\label{X40}\\
A_{2}^{(0)}(\beta) &=& {2 \over 5} \biggl[2 - {\arctan \beta \over
\beta^3} (9 + 5 \beta^{2}) + {9 \over \beta^{2}}  - \beta^{2} \ln
{\rm Rm}
\nonumber \\
& & - 2 \beta^{2} \ln \biggl({1 + \beta^{2} \over 1 + \beta^{2}
\sqrt{\rm Rm}}\biggr) \biggr] \;,
\label{X41}
\end{eqnarray}
where $\beta = \sqrt{8}  B / { B}_{\rm eq}$. For $  B \ll { B}_{\rm
eq} / 4 {\rm Rm}^{1/4} $ these functions are given by
\begin{eqnarray*}
A_{1}^{(0)}(\beta) &\sim& 2 - {1 \over 5} \beta^{2} \ln {\rm Rm}
\;,
\\
A_{2}^{(0)}(\beta) &\sim& - {2 \over 5} \beta^{2} \biggl[\ln {\rm
Rm} +  {2 \over 15}\biggr]\; .
\end{eqnarray*}
For $ { B}_{\rm eq} / 4 {\rm Rm}^{1/4} \ll  B \ll { B}_{\rm eq} / 4
$ these functions are given by
\begin{eqnarray*}
A_{1}^{(0)}(\beta) &\sim& 2 + {2 \over 5} \beta^{2} \biggl[2 \ln
\beta -  {16 \over 15} + {4 \over 7} \beta^{2} \biggr] \;,
\\
A_{2}^{(0)}(\beta) &\sim& {2 \over 5} \beta^{2} \biggl[4 \ln \beta
-  {2 \over 15} - 3 \beta^{2} \biggr] \;,
\end{eqnarray*}
and for $  B \gg {B}_{\rm eq} / 4 $ they are given by
\begin{eqnarray*}
A_{1}^{(0)}(\beta) &\sim& {\pi \over \beta} - {3 \over \beta^{2}}
\;, \quad A_{2}^{(0)}(\beta) \sim - {\pi \over \beta} + {6 \over
\beta^{2}} \; .
\end{eqnarray*}

\end{document}